\documentclass{aa}
\usepackage{graphics}
\begin{document}

\thesaurus{03(11.05.1, 11.11.1, 11.01.1, 11.06.1} 

\title{The evolution of the color gradients of early-type cluster
galaxies
\thanks{Based on observations with the NASA/ESA
               {\it Hubble Space Telescope}, obtained at the Space Telescope
               Science Institute, which is operated by AURA, Inc., under
               NASA contract NAS 5-26555.} }

\author{R.P. Saglia\inst{1}, C. Maraston\inst{1}, L. Greggio\inst{1}\inst{2}, 
R.  Bender\inst{1}, B. Ziegler\inst{3}}

\offprints{R.P. Saglia (email: saglia@usm.uni-muenchen.de)}

\institute{{Universit\"atssternwarte, Scheinerstr. 1, D-81679 M\"unchen,
             Germany} 
\and{Osservatorio Astronomico, via Ranzani 1 , I-40127 Bologna, Italy}
\and{Universit\"atssternwarte, Geismarlandstr. 11, D-37083 G\"ottingen }}

\date{Received 10.12.1999, accepted 04.07.2000}

\authorrunning{Saglia et al.}

   \maketitle 
   \markboth{Saglia et al.: Evolution of color gradients}{}
   
\begin{abstract}

We investigate the origin of color gradients in cluster early-type
galaxies to probe whether pure age or
pure metallicity gradients can explain the observed data in local and
distant ($z\approx 0.4$) samples.
We measure the surface brightness profiles of the 20 brightest
early-type galaxies of CL0949+44 (hereafter CL0949) 
at redshift $z$=0.35-0.38 from HST WF2
frames taken in the filters F555W, F675W, F814W. We determine the
color profiles $(V-R)(r)$, $(V-I)(r)$, and $(R-I)(r)$ as a function of
the radial distance $r$ in arcsec, and fit
logarithmic gradients in the range $-0.2$ to 0.1 mag per decade.
These values are similar to what is found locally for the colors $(U-B)$,
$(U-V)$, $(B-V)$ which approximately match the $(V-R)$, $(V-I)$, $(R-I)$ at
redshift $\approx 0.4$.  We analyse the results with up to date
stellar population models.  We find that passive evolution of
metallicity gradients ($\approx 0.2$ dex per radial decade) provides
a consistent explanation of the local and distant galaxies' data.
Invoking pure age gradients (with fixed metallicity) to
explain local color gradients produces too steep gradients at 
redshifts $z\approx 0.4$. Pure age gradients are
consistent with the data only  if large present day ages $\ge15$
Gyr are assumed for the galaxy centers.

\keywords{galaxies: clusters: general -- galaxies: elliptical and
lenticular, cD --  galaxies: evolution -- galaxies: formation -- 
      galaxies: fundamental parameters}   

\end{abstract}

\section{Introduction}
\label{introduction}

The efforts to understand the mechanisms of formation and evolution of
elliptical galaxies have dramatically increased in the last years. 
On the theoretical side, the classical picture of the monolithic
collapse \`a la Larson (\cite{L74}) has been increasingly questioned by
scenarios of hierarchical structure formation, where ellipticals are
formed by mergers of spirals (White \& Rees \cite{WR78}). While in the
first case the stars of the galaxies are supposed to form in an early
episode of violent and rapid formation, the semi-analytical models of
galaxy formation (Kauffmann \cite{K96}, 
Baugh et al. \cite{Betal96}) predict more extended
star formation histories, with substantial fractions of the stellar
population formed at relatively small redshifts. From an observational
point of view, evidence is growing that the formation of a fair fraction of
the stellar component of bright cluster Es must have taken place at
high redshift ($z\ge 2$) and cannot have lasted for more than 1 Gyr. This
conclusion is supported at small redshifts by the existence of a number
of correlations between the global parameters of elliptical galaxies,
all with small scatter: the Color-Magnitude (CM, Bower et
al. \cite{Betal92}), the Fundamental Plane (J\o rgensen et
al. \cite{JFK96}), the Mg-$\sigma$ (Colless et al. \cite{Cetal99}) relations. 
These hold to redshifts up to 1 (Stanford et
al. \cite{SED98}, Bender et al. \cite{BSZBBGH97}, Kelson et
al. \cite{KDFIF97}, van Dokkum et
al. \cite{DFKI98}, Bender, Ziegler \& Bruzual \cite{BZB96},
Ziegler \& Bender \cite{ZB97}, Ellis et al. \cite{Eetal97}), consistent
with passive evolution and high formation-ages. In addition, the Mg
over Fe overabundance of the stellar populations of local cluster Es
(Worthey, Faber \& Gonz\'alez \cite{WFG92}, Mehlert et al. \cite{M00b}), 
requires short star formation times
and/or a flat IMF (Thomas, Greggio, Bender \cite{TGB99}). Field ellipticals
might have larger spreads in formation ages (Gonz\'ales \cite{G93}), but
comparisons with cluster Es (Bernardi et al. \cite{Betal98}) and
the first trends observed for the evolution of the FP with redshift (Treu
et al. \cite{T99}) do not allow much freedom.

The study of the variations of the stellar populations inside
galaxies, the so-called {\it gradients}, offers an additional tool to
constrain the models of galaxy formation. Classical monolithic
collapse models with Salpeter IMF (Carlberg \cite{C84}) predict
strong metallicity gradients. The merger trees generated
by hierarchical models of galaxy formation (Lacey \& Cole
\cite{LC93}) are expected to dilute
gradients originally present in the merging galaxies and therefore give
end-products with milder gradients (White \cite{W80}), although
 detailed quantitative predictions are still lacking. 
Color gradients are known to exist in ellipticals since the pioneering
work of de Vaucouleurs (\cite{dV61}). The largest modern, CCD based,
dataset has been collected by Franx et al. (\cite{FIH89}), Peletier et
al. (\cite{PDIDC90}), Goudfrooij et al.
(\cite{GHJNDV94}), who measured surface brightness profiles in the
optical $U,~B,~V,~R,~I$ bands for a large set of local ellipticals.
These radial changes in colors have been traditionally
interpreted as changes in metallicity (Faber \cite{F77}),
arguing that line index gradients measured in local field (Davies et
al. \cite{D93}, Carollo et al. \cite{C93}, Kobayashi \& Arimoto
\cite{KA99}) and cluster (Mehlert et
al. \cite{M98}, \cite{M00a}) galaxies support this view. 
A $\approx 0.2$ dex
change in metallicity per radial decade suffices to explain the
photometric and spectroscopic data (see, for example, Peletier et al. 
\cite{PDIDC90}). However, due to the age-metallicity degeneracy
of the {\it spectra} of simple stellar populations (Worthey \cite{W94}) 
this conclusion can be questioned. De Jong (\cite{dJ96}) favors radial
variations in age to explain the color gradients observed in his
sample of local spiral galaxies. If ellipticals form from merging of
spirals, age might be the driver of color gradients in these objects
too. The combined analysis of more
metal-sensitive (i.e., Mg$_2$) and more age-sensitive (i.e.,
$H\beta$) line index profiles points to a combination of both age and
metallicity variations (Gonz\'ales \cite{G93}, 
Mehlert et al. \cite{M00b}). Finally, 
Goudfrooij \& de Jong (\cite{GdJ95}) and Wise \& Silva (\cite{WS96})
stress the importance of dust on the interpretation of color gradients
and conclude that it might add an additional source of degeneracy to
the problem.

With the resolution power of HST it is now possible to measure the
photometric properties of distant galaxies and therefore implement an
additional constraint on the stellar population models, i.e. the time
evolution, that helps breaking the degeneracies discussed above. 
Abraham et al. (\cite{AEFTG99}) study color distributions of
intermediate redshift galaxies of the Hubble Deep Field to constrain
the star formation history of the Hubble Sequence of field galaxies.
Here we determine the color gradients in the $V,~R,~I$
Johnson-Cousins filters of a sample of cluster early-type
galaxies at redshift $z\approx 0.4$ observed with HST, 
and compare them to the U, B and V
gradients observed in local Es.  By means of new, up to date models of
Simple Stellar Populations (SSP), we investigate whether pure age or
pure metallicity gradients can explain the observed colors and gradients
and their evolution with redshift. 
The structure of the paper is the
following. In Sect. \ref{cl0952} we present the new HST observations
of CL0949, the data reduction and the derived photometric profiles. In
Sect. \ref{localsample} we describe the reference sample of local cluster
galaxies. In
Sect. \ref{modeling} we describe the stellar population models and
their use to study the evolution of color gradients. Ages and
metallicities derived according to the different modeling assumptions
are presented in Sect. \ref{agesandzs}. Results are
discussed in Sect. \ref{results} and conclusions are drawn in
Sect. \ref{conclusions}.

\section{HST Observations of CL0949}
\label{cl0952}

\begin{figure}
\resizebox{\hsize}{!}{\includegraphics{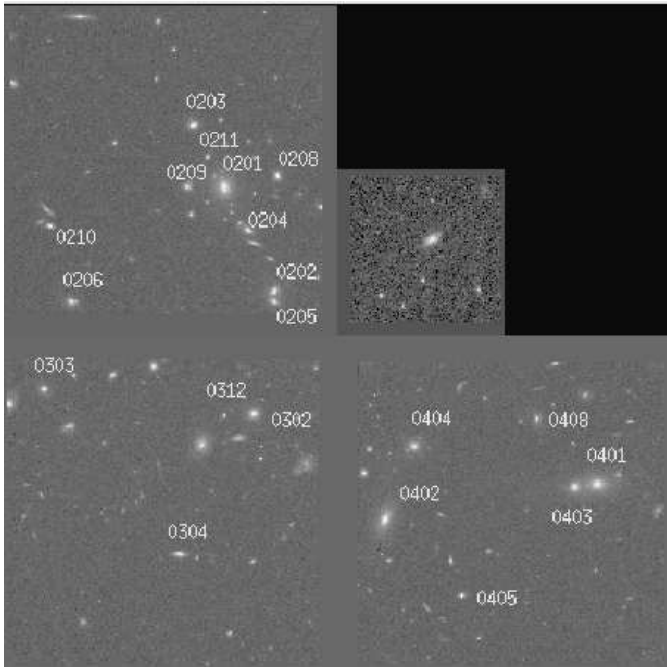}}
\caption[The F814 image of CL0949]{The F814W image of the cluster
CL0949. The labels indicate the 20 brightest early-type galaxies detected.}
\label{figcluster}
\end{figure}

The central core of CL0949 (a superposition of two clusters at
redshifts 0.35 and 0.38, Dressler \& Gunn \cite{DG92}) was observed with
HST and WFPC2 in December 1998 (PID 6478). Nine images with each of
the filters F555W, F675W, and F814W were obtained totaling exposure
times of 7200 sec, 7200 sec, and 7000 sec respectively. The images
were taken with sub-pixel shifts to improve the resolution and
sampling. The pipeline reduced frames were further processed under
IRAF to eliminate cosmic rays, and combined after resampling to half
the original scale, following the algorithm described in
Seitz et al. (\cite{SSBHBZ98}). 
The subsequent analysis was performed under MIDAS. Particular care was
needed to treat the images taken with the filter F555W, which suffered
from an elevated background level throughout the field of view, due to
Earth reflection by the optical telescope assembly and spiders, 
and the related presence of a dark diagonal X
structure. In order to determine the background accurately, the
Sextractor program (Bertin \& Arnoux \cite{BA96}) was used. 
We detected and subtracted
sources iteratively and constructed an average source-free sky
frame. The resulting source-free, sky-subtracted images taken with the
F675W and F814W filters have rms residuals of 0.3-0.5\%, the ones with
the F555W filter up to 1\%, which set the final precision of our
sky subtraction. In addition, we constructed background-subtracted
images using the first-iteration background determined by Sextractor,
which takes into account the diffuse light contribution of extended
objects. These frames were then used to perform the surface photometry 
analysis for all galaxies except for the cD on the WF2 frame (\#201 of 
Fig.~\ref{figcluster}). In this case we preferred the sky-subtracted images
to be able to measure the extended halo of the object. 

The isophote shape analysis was performed following Bender and
M\"ollenhoff (\cite{BM87}).  The
procedure was improved in two aspects. 1) Mask files were produced
automatically from the ``segmentation frames'' produced by
Sextractor. 2) The surface brightness and integrated magnitude profiles
following elliptical isophotes were computed from the averages of the
not masked pixels between the isophotes. Statistical errors were also
derived from the measured rms. Profiles of surface brightness, ellipticity,
position angle and higher order terms of the Fourier
decomposition were derived for the 70 brightest objects detected by
Sextractor on the F814W frames. The information was used in
combination with the visual inspection of the images to classify the
galaxies morphologically. Twenty early-type galaxies were found. Their
identification numbers can be read in Fig. \ref{figcluster}. In the
following we shall consider only these objects. The data for the
late-type galaxies will be presented elsewhere. The cross-identification
with Dressler \& Gunn (\cite{DG92}) is given in Table \ref{tabident},
where the available spectroscopic information is also summarized.
Inspection of the profiles of the isophote shape parameters
shows smooth variations as a function of the radial distance, 
and absence of strong odd-term coefficients,
which
excludes the presence of large amounts of patchy dust. This is
confirmed by the examination of the two-dimensional (V-R), (V-I),
(R-I) color maps.

\begin{table*}
\caption[Cross-identification]{Results of the photometric fits. Column
1 gives the galaxy name (see Fig. \ref{figcluster}), Col. 2 the 
cross identification with Dressler \& Gunn (\cite{DG92}), Col. 3 the
redshift from Dressler \& Gunn (\cite{DG92}), Col. 4 the morphological
type, Cols. 5-10 give the half-luminosity radii $R_e$ in arcsec and 
total magnitudes $m_T$ 
in the F555W, F675W, and F814W filters respectively. Magnitudes are
calibrated to the $V,~R,~I$ Johnson-Cousins system (see text). 
Col. 11 gives the central
velocity dispersions from Ziegler \& Bender (\cite{ZB97}) and comments.}
\begin{flushleft}
\begin{tabular}{rrrrrrrrrrr}
\noalign{\smallskip}
\hline
\noalign{\smallskip}
Name &	D\&G92	& z	& Type & $R_e(555)$ & $m_T(555)$ &$R_e(675)$ & $m_T(675)$ & $R_e(814)$ & $m_T(814)$ & Comment\\
\noalign{\smallskip}
\hline
201 & 118 &  .3815 & cD	    & 1.46 &   19.44  &   1.90$^{1}$  &  18.06$^{1}$  &  2.05$^{1}$  & 17.23$^{1}$ &   $\sigma=325\pm50$\\
202 &     &        & E	    & 0.64  &  20.54   &  0.59   & 19.53   & 0.55   & 18.78 & 		   \\
203 & 102 &  .3766 & SB0    & 0.35  &  20.52   &  0.37   & 19.58   & 0.36   & 18.87 &  E+A \\
204 & 130 &  .3402 & E	    & 0.63  &  20.77   &  0.58   & 19.85   & 0.55   & 19.09 & 		   \\
205 &     &        & E/S0   & 0.52 &   20.91  &   0.49  &  19.86  &  0.48  & 19.08 & 		   \\
206 &     &        & E	    & 0.72 &   20.84  &   0.58  &  19.91  &  0.54  & 19.17 & 		   \\
208 & 123 &  .3779 & S0	    & 0.45 &   21.51  &   0.47  &   20.4  &  0.50  & 19.58 & 		   \\
209 & 110 &  .3744 & S/S0   & 0.67  &   21.3   &  0.66   & 20.34   & 0.62   & 19.74 & 		   \\
210 &     &        & E/S0   & 0.44 &   21.45  &   0.34  &  20.51  &  0.41  & 19.69 & 		   \\
211 & 108 &  .3799 & E/S0   & 0.20 &   21.87  &   0.18  &  20.79  &  0.20  & 20.03 & 		   \\
302 & 163 &  .3778 & E/S0   & 0.65 &   20.42  &   0.58  &  19.47  &  0.57  & 18.70 & 		   \\
303 & 119 &  .3484 & E/S0   & 0.43 &      21  &   0.49  &  19.88  &  0.44  & 19.17 & 		   \\
304 &     &        & S0	    & 0.56 &    21.1  &   0.47  &  20.11  &  0.48  & 19.31 & 		   \\
312 &     &        & E/S0   & 0.15 &    22.5  &   0.16  &  21.39  &  0.15  & 20.67 & 		   \\
401 & 221 &  .3772 & E	    & 1.07  &  20.05   &   1.05   & 18.87   & 1.07   & 18.06 &  $\sigma=230\pm25$	\\
402 & 193 &  .3767 & Sa/S0  & 1.12 &      20  &   1.10  &  18.93  &   1.12  & 18.14 &  $\sigma=230\pm25$	\\
403 & 217 &  .3784 & E	    & 0.69 &   20.42  &   0.71  &  19.32  &  0.69  & 18.57 &  $\sigma=125\pm15$	\\
404 & 187 &  .3778 & E	    & 0.81 &   20.63  &   0.77  &  19.61  &  0.76  & 18.87 & 		   \\
405 &     &        & SB0    & 0.37  &  21.79   &  0.29   & 20.83   & 0.29   & 20.05 & 		   \\
408 &     &        & S0     & $-^{2}$ &   $-^{2}$  &   1.03  &   20.7  & 0.85  & 19.78 & Background?   \\
\noalign{\smallskip}
\hline
\multicolumn{11}{l}{1: uncertain}\\
\multicolumn{11}{l}{2: profile too short to allow a reliable fit}\\
\end{tabular}
\end{flushleft}
\label{tabident}
\end{table*}

Circularly averaged surface brightness profiles were also measured
following Saglia et al. (\cite{SBBBCDMW97a}). The photometric
calibration on the $V,~R,~I$ Johnson-Cousins bands was performed
using the ``synthetic system'' (Holzmann et al. \cite{HBCHTWW95}) 
as in Fig. 1 of
Ziegler et al. (\cite{ZSBBGS99}).  
A first estimate of the zero-points was obtained
using the $(V-I)$ and $(V-R)$ colors typical of $z=0.4$ early-type
galaxies, as in Ziegler et al. (\cite{ZSBBGS99}). 
The final zero-points were derived
recomputing the calibration with the so-measured $(V-I)$ and $(V-R)$
colors.  Half-luminosity radii $R_e(555)$, $R_e(675)$, $R_e(814)$ and
total magnitudes $m_T(555)$, $m_T(675)$, $m_T(814)$ were derived fitting the
circularly averaged profiles using the algorithm described and
extensively tested in Saglia et al. (\cite{SBBBCDMW97b}). 
The algorithm fits a PSF (computed using the Tinytim program)
broadened $r^{1/4}$ and an
exponential component simultaneously and separately to the circularly
averaged surface brightness profiles. As a result, disk-to-bulge
ratios and the scale-lengths of the bulge and the disk components are
also determined. The quality of the fits were
explored by Monte Carlo simulations in Saglia et
al. (\cite{SBBBCDMW97b}), taking into account the possible influence of
sky subtraction errors, the signal-to-noise ratio, the
radial extent of the profiles and the $\chi^2$ quality of the fit.
Accordingly, the half-luminosity radii are expected to be accurate to
25\%, the total magnitudes to 0.15 mag. 
Two fits are more uncertain than these figures. The values of
$R_e(814)$ and $R_e(675)$ of galaxy 201 have been derived restricting
the fit to the radial range of the F555W profile. 
Fitting the whole F814W profile, more
extended than the F555W and F675W ones and shallower than the
extrapolated profiles expected from the fits in the V and R bands, one
would get a value of $R_e(814)$ twice as large. Fitting the whole
F675W profile produces a value of $R_e(675)$ 30\% smaller.
Finally, the F555W profile of the galaxy 408 is not extended enough to allow a
reliable determination  of $R_e(555)$. 
The best fit the these data involves a large (50\%) extrapolation and
gives  a value of $R_e(555)$ twice as  large as $R_e(675)$ and
$R_e(814)$. 

\begin{figure*}
\resizebox{\hsize}{!}{\includegraphics{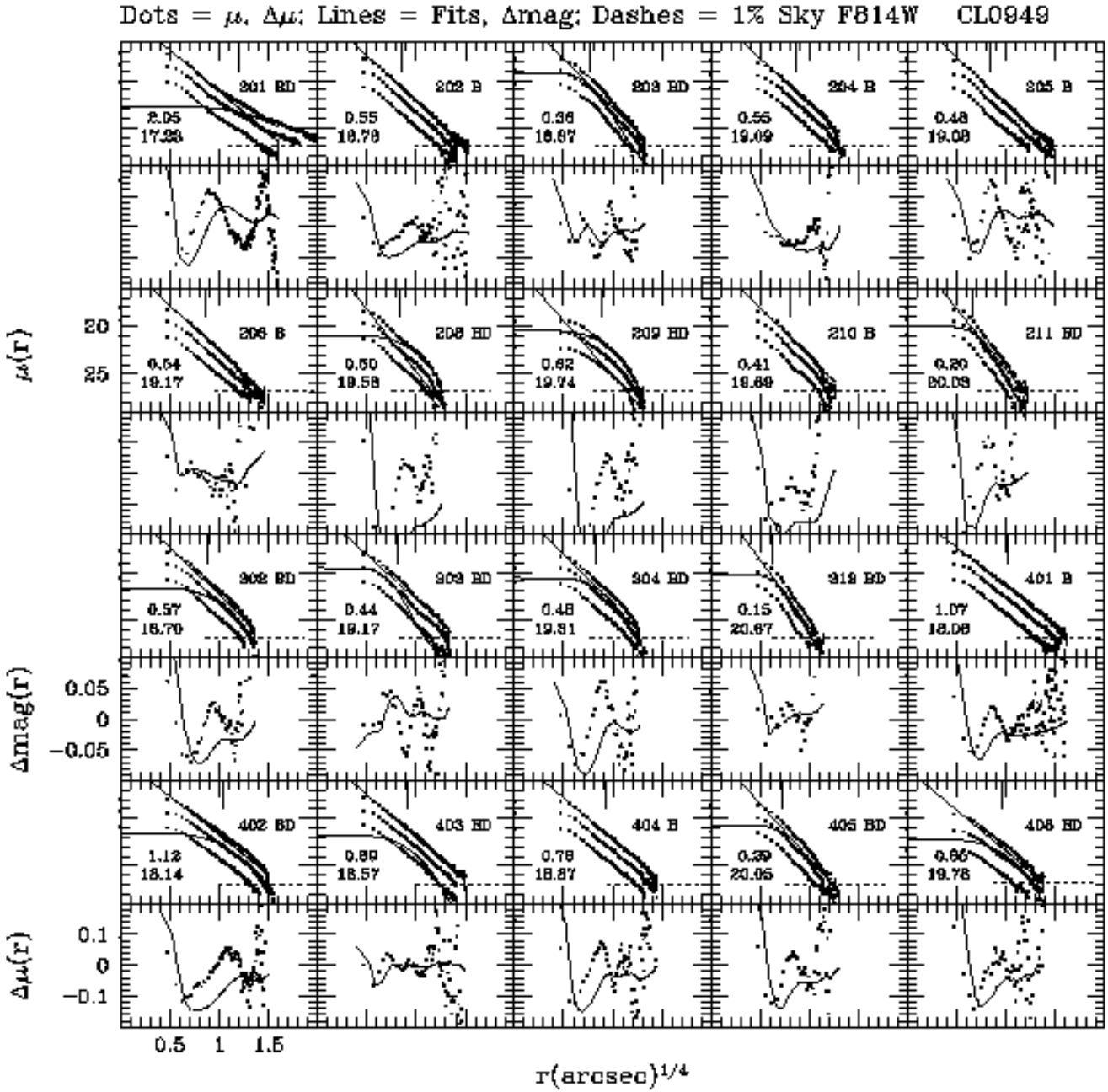}}
\caption[The fits to the surface brightness profiles]{The fits to the
circularly averaged surface brightness profiles (taken with the
F814W filter and calibrated in the Cousins I band) 
of the 20 brightest early-type galaxies of CL0949. Two
plots are shown for each galaxy. The upper panel shows the calibrated
surface brightness profile (small dots) together with the fitted bulge
and/or disk profiles (full lines) as a function of the 1/4 power of
the distance $r$ from the center in arcsec. 
The horizontal dashed line shows
the position corresponding to 1 per cent of the sky level.
We also show the circularly averaged surface brightness profiles
taken with the F675W and F555W filters, calibrated in the Johnson-Cousins V
and R bands. For clarity they have been shifted by 0.5 and 1 mag respectively.
 In each panel
the galaxy name and the type of best fit are
given (B for bulge only, D for disk only, BD for bulge+disk fit),
together with the value of $R_e$ in arcsec and
the total magnitude calibrated in the Cousins I band. 
The $R_e$ location is also marked as a vertical
line from the top xaxis in each panel. Below each surface brightness profile
we show the residuals of the fits, $\Delta \mu(r)=\mu_{obs}(r)-\mu_{fit}(r)$
(dots, scale -0.2 to 0.2 mag arcsec$^{-2}$) and the integrated
magnitude difference $\Delta mag(r)=mag_{obs}(<r)-mag_{fit}(<r)$ (full
lines, scale -0.1 to 0.1 mag). }
\label{figfits}
\end{figure*}

Fig. \ref{figfits} shows the resulting fits to the F814W filter
profiles and the profiles in the F675W and F555W bands. 

Half-luminosity radii and total magnitudes are given
in Table \ref{tabident}. The scales ($R_{eB}$ and $h$ in arcsec) and the
magnitudes ($m_B$ and $m_D$) of the bulge and
disk components measured with the F814W profiles 
are given in Table \ref{tabbulgedisk} together with
the $D/B$ ratio, for the objects where a two-component model was fit. 
As described above, the fit to the galaxy 201 is highly uncertain. If
the whole radial extent of the F814W profile is fit, an 8 times larger
value of the bulge scalelength is needed. 

Following Burstein \& Heiles (\cite{BH84}), no correction for galactic 
absorption was necessary. Note, however, that Schlegel et
al. (\cite{SFD98}) give $E(B-V)=0.019$. The use of this correction
gives $A_V=0.063$, $A_R=0.051$, $A_I=0.037$, making the derived intrinsic
$(V-R)$, $(V-I)$, $(R-I)$ colors bluer by 0.01, 0.026, 0.014 mag
respectively. These variations do not affect what follows.

\begin{table}
\caption[Bulge+Disk]{The parameters of the bulge+disk fits as measured
with the F814W profiles. Col. 1 gives the galaxy name (see
Fig. \ref{figcluster}), Col. 2 the half-light radius of the bulge component 
$R_{eB}$ in arcsec, Col. 3 the magnitude of the bulge component
$m_B$, Col. 4 the scalelength $h$ of the disk component in arcsec,
Col. 5 the magnitude of the disk component $m_D$, Col. 6 the
disk-to-bulge ratio $D/B$. Magnitudes are calibrated to the I Cousins
filter (see text).}
\begin{flushleft}
\begin{tabular}{rrrrrr}
\noalign{\smallskip}
\hline
\noalign{\smallskip}
Name & $R_{eB}(814)$ & $m_B(814)$ & $h(814)$ & $m_D(814)$ & $D/B$ \\
\noalign{\smallskip}
\hline
         201$^{1}$ &0.80  &     17.81  &     3.08 &      18.18 &     0.71\\
         203  &	    0.44  &     19.32  &     0.18 &      20.05 &     0.51\\
         208  &	    0.16  &      20.5  &     0.48 &      20.18 &      1.34\\
         209  &	    0.26  &     21.42  &     0.42 &         20 &      3.69\\
         211  &	    0.18  &      20.4  &     0.14 &      21.36 &     0.42\\
         302  &	    0.49  &     18.89  &     0.54 &      20.68 &     0.19\\
         303  &	    0.60  &     19.41  &     0.13 &      20.92 &       0.25\\
         304  &	    0.38  &     19.81  &     0.36 &      20.41 &     0.58\\
         312  &	    0.12  &     21.13  &     0.11 &      21.83 &     0.53\\
         402  &	    1.02  &      18.4  &     0.80 &      19.82 &      0.27\\
         403  &	    0.67  &     18.67  &     0.45 &      21.21 &    0.097\\
         405  &	    0.25  &     20.48  &     0.19 &      21.26 &     0.49\\
         408  &	    0.82  &     20.09  &     0.54 &       21.3 &     0.33\\
\noalign{\smallskip}
\hline
\multicolumn{6}{l}{1: uncertain fit}\\
\end{tabular}
\end{flushleft}
\label{tabbulgedisk}
\end{table}

\begin{figure*}[ht!]
\resizebox{\hsize}{!}{\includegraphics{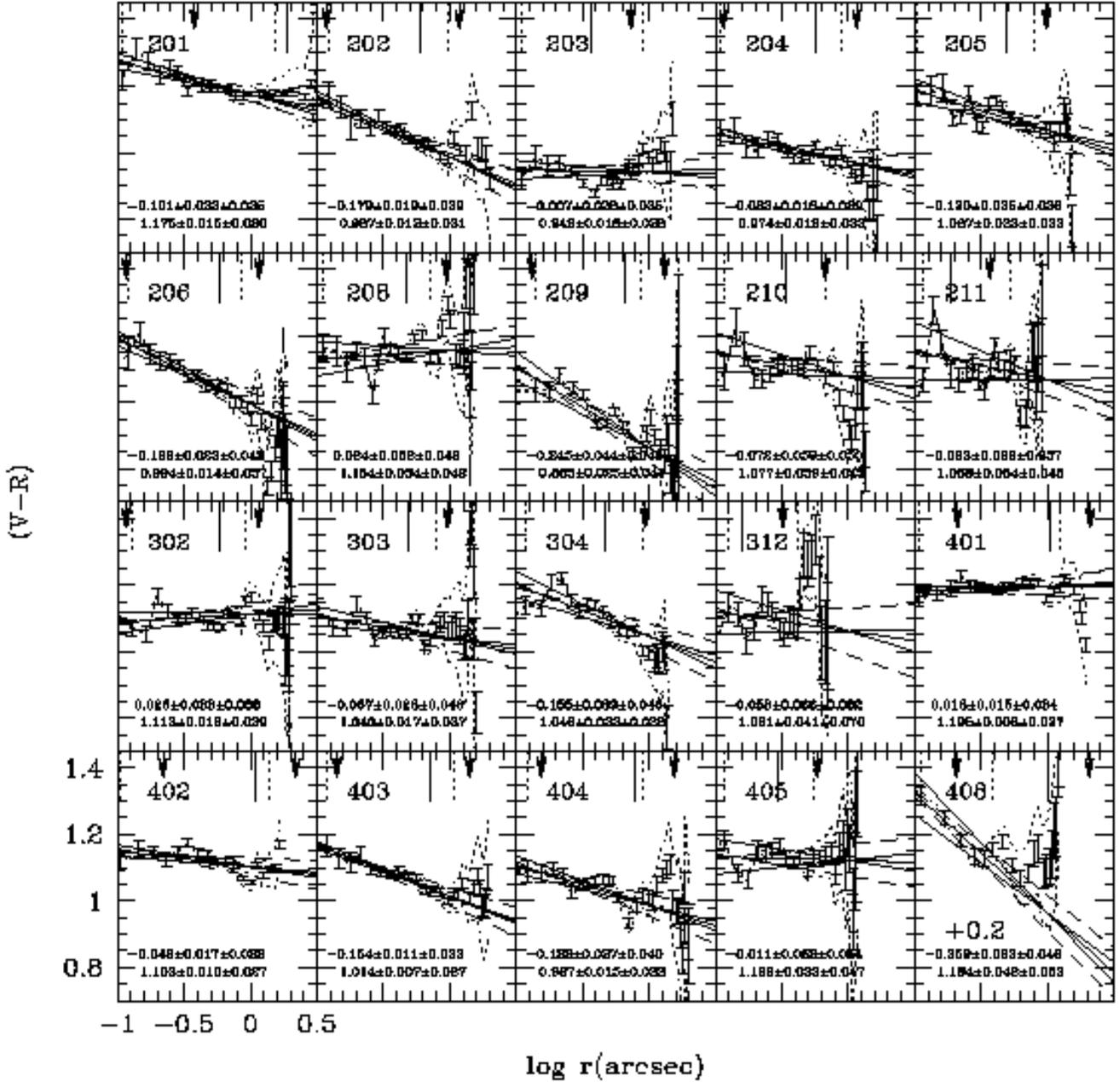}}
\caption[Color gradients]{The F555W-F675W color profiles (with
statistical errors) and the logarithmic gradient fits 
of the 20 brightest early-type galaxies of CL0949. Notice that since
the colors of N408 are particularly red, we plot the observed colors
shifted by -0.2 mag. The shift of 0.2 mag indicated 
in the figure has to be added to get the real colors of the object.
The dotted curved lines show the systematic errors on the colors 
due to sky subtraction. The straight full
lines show the fitted gradients and their statistical errors. The
straight dashed lines show their systematic errors. Relevant radial
distances are marked on the top x axis of each panel as vertical
lines: the location of  $R_e(675)$ (solid); the minimum and maximum
radius used in the fit (dotted); the $0.2 R_e-2R_e$ range (arrows) used in the
modeling. Finally, each panel is labeled with 
the galaxy name, and with the gradient slope (upper row) and zero
point (lower row) followed by their statistical and systematic error.}
\label{figgrad555675}
\end{figure*}

\begin{figure*}[ht!]
\resizebox{\hsize}{!}{\includegraphics{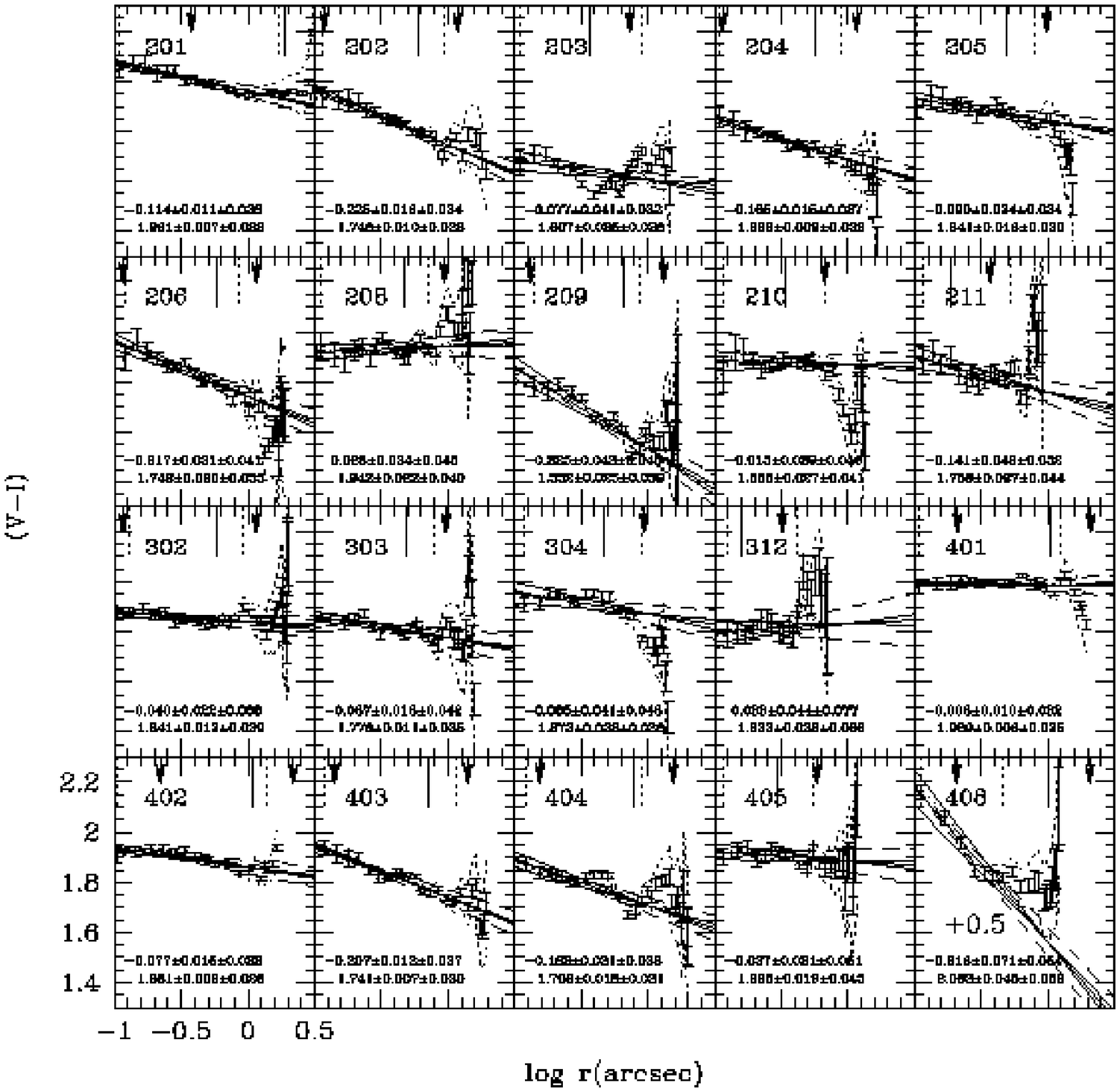}}
\caption[Color gradients]{The F555W-F814W color profiles and
logarithmic gradient fits of the 20 brightest early-type galaxies of
CL0949. Lines and labels as in Fig. \ref{figgrad555675}. The shift
to be applied to the colors of N408 is shown as in Fig. \ref{figgrad555675}.}
\label{figgrad555814}
\end{figure*}

\begin{figure*}[ht!]
\resizebox{\hsize}{!}{\includegraphics{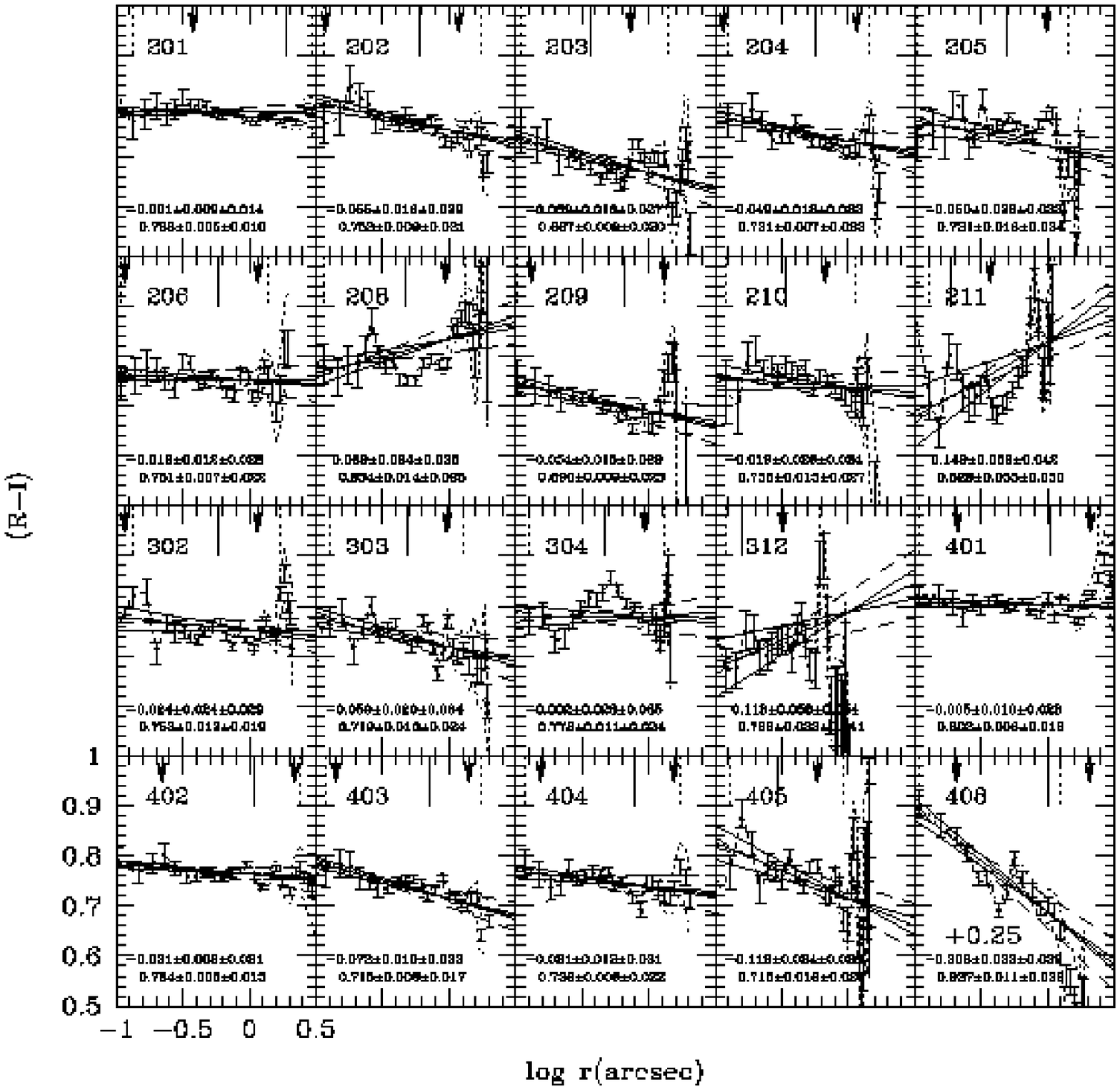}}
\caption[Color gradients]{The F675W-F814W color profiles and
logarithmic gradient fits of the 20 brightest early-type galaxies of CL0949. 
Lines and labels as in Fig. \ref{figgrad555675}. The shift
to be applied to the colors of N408 is shown as in Fig. \ref{figgrad555675}.}
\label{figgrad675814}
\end{figure*}

Finally, we determined the color profiles of the galaxies from the
calibrated elliptical isophote surface brightness  profiles, estimating
statistical and systematic errors. The latter were computed
considering the extreme cases of possible sky subtraction errors, $\pm
0.5$\% for the F675W and F814W filters, and $\pm 1$\% for the F555W
(see above). For example, we computed the two $(V-I)$ profiles
corresponding to +1\% sky error on V and -0.5\% sky error on I, and
-1\% sky error on V and +0.5\% sky error on I. The error budget on the
colors is dominated by statistics in the central regions of the
galaxies, and by systematics in the outer parts. We fitted the
logarithmic slope gradient and zero point, weighting the data
according to the statistical errors. In the fit we consider only
points within  0.1 ($R_{min}$)and 3.2 ($R_{max}$) arcsec and with
statistical and systematic errors less than 0.1 mag. Systematic errors on the
gradients and zero-points due to sky subtraction were
estimated determining the two quantities for the two extreme cases of
possible sky errors discussed above. 
The median value of $R_{min}/R_e$ is 0.2, of $R_{max}/R_e$ 1.6 for the
$(V-R)$ and $(V-I)$ colors, and 2.8 for $(R-I)$. We explored the
effect of the radial range by repeating the fits on a fractional
($0.2\le R/R_e\le 1.5$) or  on a fixed radial range ($0.1<R<1$
arcsec). In both cases we obtained similar gradients and zeropoints 
within the statistical
errors, if the systematic errors in this radial range due to sky subtraction 
are not too large ($\approx 0.1$ mag). 
We verified that 
the differences between the HST PSFs in the different filters do not
affect the derived gradients within the errors. 
Tests on $R^{1/4}$ profiles of $R_e=0.5$ arcsec convolved with the PSF
for the filters F555W and F675W show that the systematic errors
produced on the logarithmic gradients computed following our procedure
are smaller than the systematics due to the sky subtraction errors.
For galaxy 312, the smallest of the sample ($R_e\approx 0.15''$), 
we derived color gradients
by first convolving the pairs of images (for example, F675W and F555W)
with the corresponding interchanged PSFs (F555W and F675W). The
differences in the derived logarithmic slopes are smaller than twice
the statistical error; 
thus we expect that this effect is negligible
for the other, more extended objects.
Figs. \ref{figgrad555675} to 
\ref{figgrad675814} show the color profiles with the logarithmic gradient fits.
Tables \ref{tabresultsvr}-\ref{tabresultsri} summarize the results of
these fits. The colors given there are computed at half $R_e$ in the
F675W band. The galaxy 408 has colors 0.2-0.5 mag redder than the rest
of the sample. We assume that it is a background object and do not
consider it in the following analysis.
Inspection of Figs. \ref{figgrad555675} to \ref{figgrad675814} shows
that on the whole the color profiles are well described by the
logarithmic fits, with mean rms residuals in the range 0.02-0.03
mag. However, the fits are not optimal in a statistical sense, because
their reduced $\chi^2$ are rather large, in the range 2-4. This stems
from the small radial scale color variations present in many
profiles. Similar effects are observed in local galaxies (i.g.,
Peletier et al. \cite{PDIDC90}).  

\begin{table}[ht]
\caption[results]{The results of the fits to the (V-R) color
profiles. Col. 1 gives the galaxy name (see Fig. \ref{figcluster}),
Col. 2 the fitted logarithmic gradient in mag per radial decade,
Col. 3 the statistical error, Col. 4 the systematic error, Col. 5 the
fitted color at 0.5 $R_e(675)$, Col. 6 the statistical error, Col. 7
the systematic error.}
\begin{flushleft}
\begin{tabular}{rrrrrrr}
\noalign{\smallskip}
\hline
\noalign{\smallskip}
Name & a(V-R)&da$_{st}$&da$_{sy}$&Col&dCol$_{st}$&dCol$_{sy}$\\ 
\hline
 201 & -0.101 &  0.023 &  0.025 &  1.178 &  0.015 &  0.019  \\
 202 & -0.179 &  0.019 &  0.039 &  1.082 &  0.016 &  0.011  \\
 203 & -0.007 &  0.028 &  0.035 &  0.947 &  0.027 &  0.002  \\
 204 & -0.083 &  0.018 &  0.039 &  1.019 &  0.015 &  0.012  \\
 205 & -0.120 &  0.035 &  0.036 &  1.141 &  0.031 &  0.009  \\
 206 & -0.198 &  0.023 &  0.043 &  1.100 &  0.019 &  0.014  \\
 208 &  0.024 &  0.052 &  0.049 &  1.139 &  0.047 &  0.012  \\
 209 & -0.245 &  0.044 &  0.048 &  0.981 &  0.033 &  0.018  \\
 210 & -0.072 &  0.059 &  0.050 &  1.134 &  0.060 &  0.005  \\
 211 & -0.083 &  0.088 &  0.057 &  1.155 &  0.106 &  0.012  \\
 302 &  0.026 &  0.033 &  0.038 &  1.099 &  0.026 &  0.009  \\
 303 & -0.067 &  0.028 &  0.043 &  1.082 &  0.024 &  0.010  \\
 304 & -0.155 &  0.039 &  0.046 &  1.144 &  0.033 &  0.009  \\
 312 & -0.058 &  0.066 &  0.082 &  1.125 &  0.083 &  0.020  \\
 401 &  0.013 &  0.015 &  0.034 &  1.192 &  0.009 &  0.017  \\
 402 & -0.048 &  0.017 &  0.033 &  1.115 &  0.011 &  0.019  \\
 403 & -0.154 &  0.011 &  0.033 &  1.084 &  0.009 &  0.012  \\
 404 & -0.123 &  0.027 &  0.040 &  1.037 &  0.019 &  0.016  \\
 405 & -0.011 &  0.053 &  0.054 &  1.131 &  0.055 &  0.002  \\
 408 & -0.359 &  0.063 &  0.046 &  1.268 &  0.046 &  0.040  \\
\hline
\end{tabular}
\end{flushleft}
\label{tabresultsvr}
\end{table}

\begin{table}[ht]
\caption[results]{The results of the fits to the (V-I) color profiles.
Columns as in Table \ref{tabresultsvr}.}
\begin{flushleft}
\begin{tabular}{rrrrrrr}
\noalign{\smallskip}
\hline
\noalign{\smallskip}
Name & a(V-I)&da$_{st}$&da$_{sy}$&Col&dCol$_{st}$&dCol$_{sy}$\\
\hline
 201 & -0.114 &  0.011 &  0.028 &  1.963 &  0.007 &  0.021  \\
 202 & -0.225 &  0.016 &  0.034 &  1.864 &  0.013 &  0.010  \\
 203 & -0.077 &  0.041 &  0.032 &  1.663 &  0.039 &  0.003  \\
 204 & -0.165 &  0.015 &  0.037 &  1.777 &  0.012 &  0.012  \\
 205 & -0.090 &  0.024 &  0.034 &  1.896 &  0.022 &  0.009  \\
 206 & -0.217 &  0.031 &  0.041 &  1.858 &  0.026 &  0.013  \\
 208 &  0.026 &  0.034 &  0.046 &  1.926 &  0.031 &  0.011  \\
 209 & -0.325 &  0.043 &  0.045 &  1.688 &  0.033 &  0.017  \\
 210 & -0.015 &  0.039 &  0.046 &  1.879 &  0.041 &  0.005  \\
 211 & -0.141 &  0.042 &  0.052 &  1.904 &  0.051 &  0.010  \\
 302 & -0.040 &  0.022 &  0.038 &  1.862 &  0.016 &  0.008  \\
 303 & -0.087 &  0.018 &  0.042 &  1.830 &  0.016 &  0.010  \\
 304 & -0.085 &  0.041 &  0.043 &  1.927 &  0.035 &  0.008  \\
 312 &  0.032 &  0.044 &  0.077 &  1.797 &  0.055 &  0.018  \\
 401 & -0.008 &  0.010 &  0.032 &  1.991 &  0.007 &  0.016  \\
 402 & -0.077 &  0.015 &  0.032 &  1.881 &  0.009 &  0.018  \\
 403 & -0.207 &  0.012 &  0.037 &  1.835 &  0.009 &  0.013  \\
 404 & -0.182 &  0.031 &  0.038 &  1.783 &  0.022 &  0.015  \\
 405 & -0.037 &  0.031 &  0.051 &  1.917 &  0.032 &  0.002  \\
 408 & -0.616 &  0.071 &  0.054 &  2.241 &  0.050 &  0.043  \\  
\hline
\end{tabular}
\end{flushleft}
\label{tabresultsvi}
\end{table}

\begin{table}[ht]
\caption[results]{The results of the fits to the (R-I) color profiles.
Columns as in Table \ref{tabresultsvr}.}
\begin{flushleft}
\begin{tabular}{rrrrrrr}
\noalign{\smallskip}
\hline
\noalign{\smallskip}
Name & a(R-I)&da$_{st}$&da$_{sy}$&Col&dCol$_{st}$&dCol$_{sy}$\\
\hline
 201 & -0.001 &  0.009 &  0.014 &  0.788 &  0.005 &  0.009  \\
 202 & -0.055 &  0.016 &  0.029 &  0.782 &  0.012 &  0.005  \\
 203 & -0.069 &  0.016 &  0.027 &  0.718 &  0.015 &  0.000  \\
 204 & -0.049 &  0.013 &  0.032 &  0.757 &  0.010 &  0.005  \\
 205 & -0.050 &  0.028 &  0.033 &  0.751 &  0.023 &  0.004  \\
 206 & -0.012 &  0.012 &  0.029 &  0.758 &  0.010 &  0.006  \\
 208 &  0.069 &  0.024 &  0.035 &  0.790 &  0.020 &  0.004  \\
 209 & -0.054 &  0.015 &  0.029 &  0.716 &  0.011 &  0.009  \\
 210 & -0.019 &  0.026 &  0.034 &  0.753 &  0.025 &  0.000  \\
 211 &  0.149 &  0.059 &  0.042 &  0.673 &  0.070 &  0.013  \\
 302 & -0.024 &  0.024 &  0.029 &  0.766 &  0.018 &  0.003  \\
 303 & -0.059 &  0.020 &  0.034 &  0.755 &  0.016 &  0.003  \\
 304 & -0.002 &  0.023 &  0.035 &  0.779 &  0.018 &  0.002  \\
 312 &  0.113 &  0.058 &  0.054 &  0.664 &  0.072 &  0.018  \\
 401 & -0.005 &  0.010 &  0.023 &  0.804 &  0.006 &  0.009  \\
 402 & -0.021 &  0.008 &  0.021 &  0.769 &  0.005 &  0.010  \\
 403 & -0.072 &  0.010 &  0.023 &  0.747 &  0.007 &  0.006  \\
 404 & -0.031 &  0.012 &  0.031 &  0.751 &  0.008 &  0.009  \\
 405 & -0.112 &  0.034 &  0.038 &  0.808 &  0.034 &  0.004  \\
 408 & -0.208 &  0.022 &  0.039 &  0.999 &  0.013 &  0.020  \\
\hline
\end{tabular}
\end{flushleft}
\label{tabresultsri}
\end{table}

\begin{figure}[ht!]
\resizebox{\hsize}{!}{\includegraphics{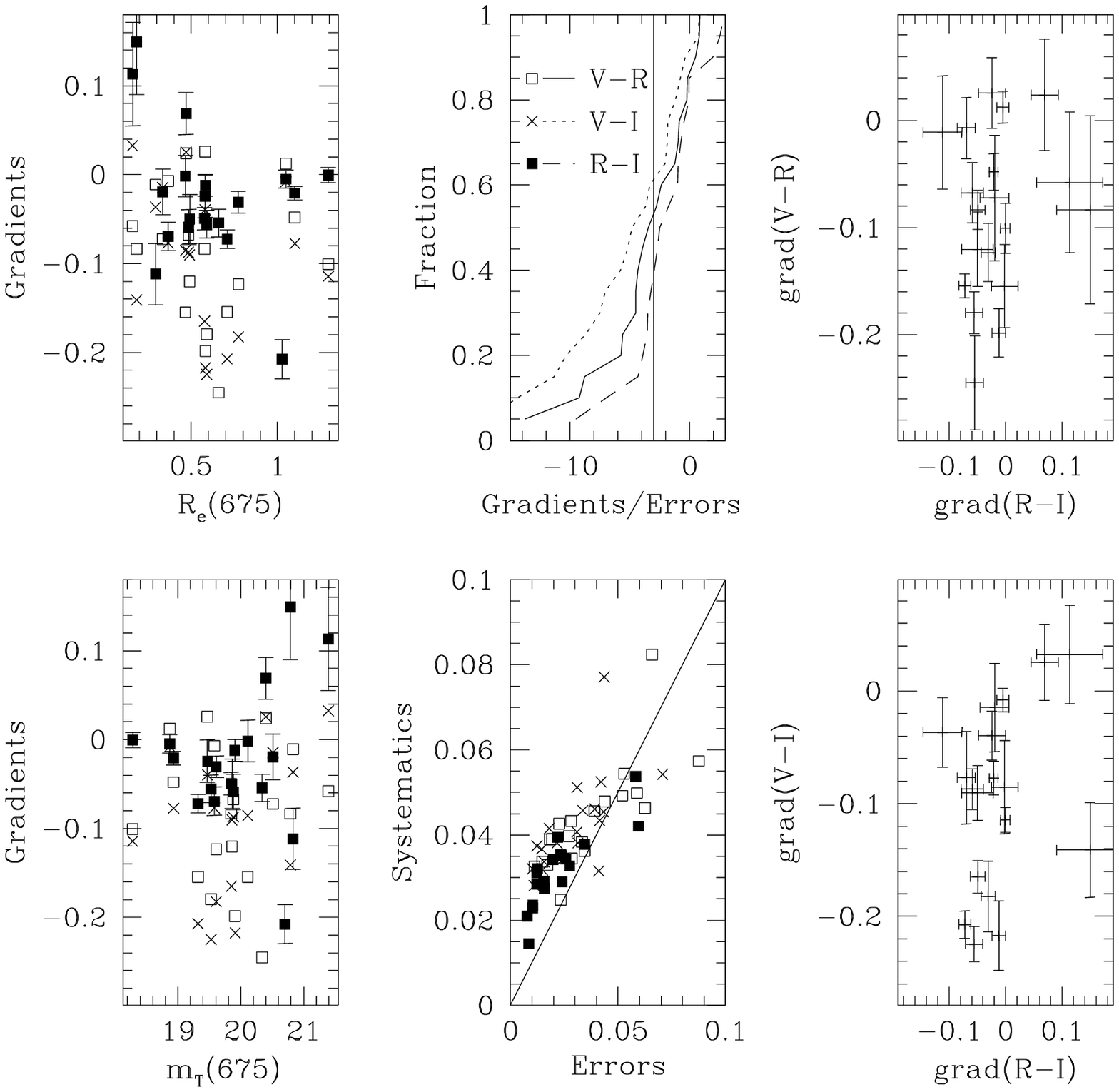}}
\caption[Errors on color gradients]{Top-left: the color gradients of
the distant sample as a
function of $R_e(675)$ in arcsec. Open squares refer to the $(V-R)$,
crosses to the $(V-I)$ and filled squares to the $(R-I)$ colors. The
bars show the statistical errors. Bottom-left: the color gradients of
the distant sample as a function of $m_T(675)$. The bars show the
systematic errors. Top-middle: the cumulative distribution of ratio
between the gradient and the statistical error. 
The solid line refers to the
$(V-R)$, the dotted to the $(V-I)$, the dashed to the $(R-I)$
colors. The vertical line shows the $3\sigma$ limit. Bottom-middle: the
correlation between statistical and systematic errors. Symbols as
above. Top-right: the correlation between the $(V-R)$ and the $(R-I)$
gradients. Statistical errors are shown. Bottom-right: the correlation 
between the $(V-I)$ and the $(R-I)$
gradients. Statistical errors are shown.}
\label{figerrgrad}
\end{figure}

Fig. \ref{figerrgrad} summarizes the properties of the gradient
sample. No correlation is observed between galaxy size and/or galaxy
magnitude and gradients. All objects have detected negative gradients in 
at least one color. For at least half of the sample the
gradients are detected with at least $3\sigma$ significance. None of
the positive gradients is statistically significant. The systematic
errors are on average larger than the statistical ones.  
Due to the rather large errors, the correlation between the
gradients in the different bands is not very strong. 

\section{The Local Sample}
\label{localsample}

\begin{figure}[ht!]
\resizebox{\hsize}{!}{\includegraphics{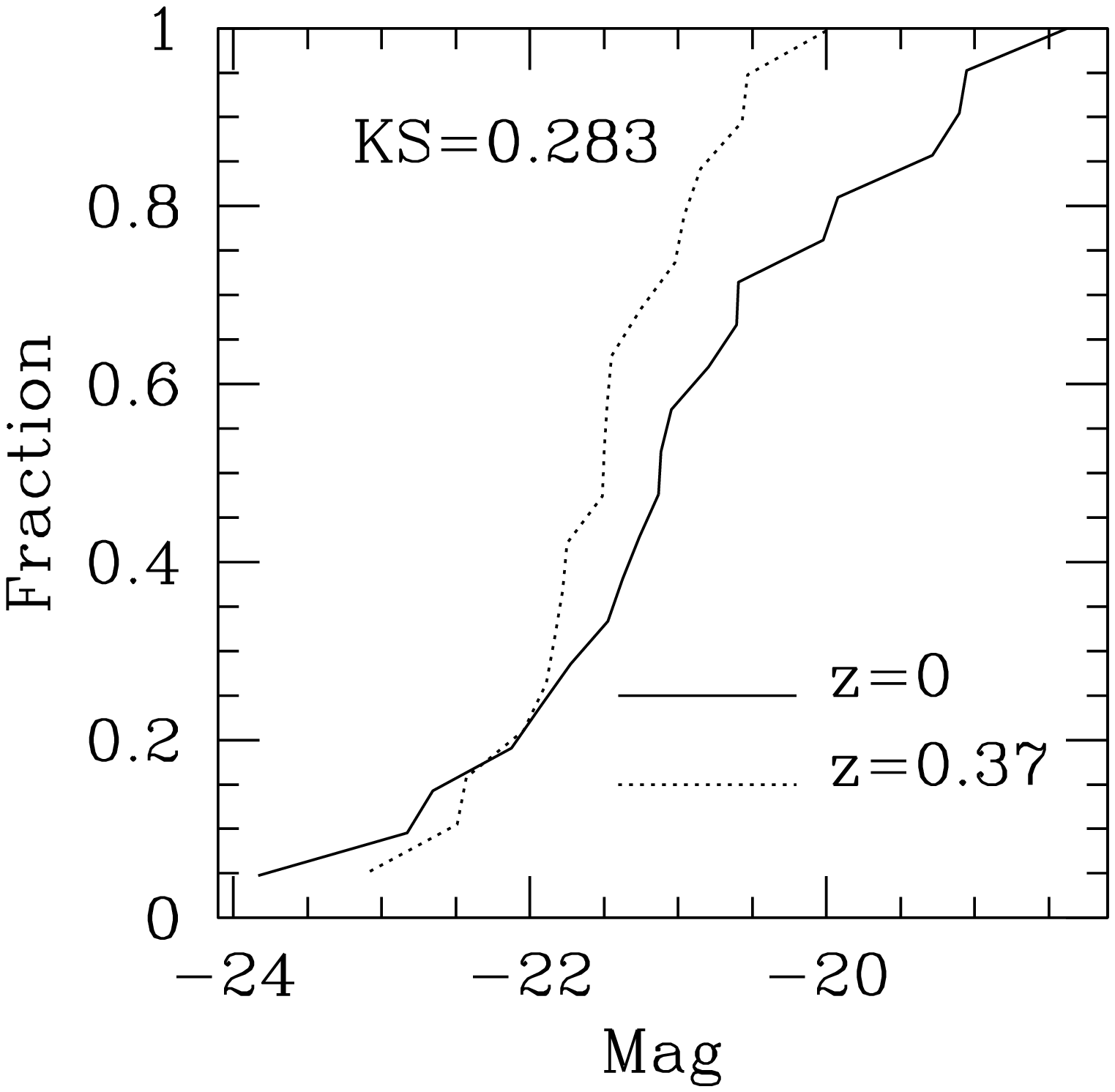}}
\caption[Cumulative distributions of magnitudes]
{Cumulative distributions of the B-band absolute magnitudes of the z=0 
(solid) and $z\approx 0.4$ (dotted) samples.}
\label{figmag}
\end{figure}

Since the F555W, F675W, F814W bands at $z\approx 0.4$ map
approximately the $U$, $B$, $V$ Johnson filters at $z=0$, the
direct comparison of our measured color gradients with those derived
for local early-type galaxies in the U, B, V bands can give a first
idea of their evolution. Therefore, we consider the surface brightness 
profiles measured by Franx et al. (\cite{FIH89}), 
Peletier et al. (\cite{PDIDC90}), and Goudfrooij et al. (\cite{GHJNDV94}). 
Focusing on cluster galaxies our sample consists of 21 
objects, for which at least one color gradient is available.
Table \ref{tablocal} lists the selected cluster galaxies, their
membership, the color gradient, the colors at $R_e/2$ and the source of the photometry. 
For those objects with source of photometry ``I'', the $V$ profile was not
available. We derived it by interpolation of the B and R profiles
using the relation $(B-V) = 0.66\times(B-R)-0.03$, which was determined from the SSP
models of Maraston (2000) for ages older than 4 Gyr, independently
of the metallicities. The errors on the
quantities given
in Table \ref{tablocal} are systematic and due to the uncertainties in
the sky subtraction. The statistical errors are (usually) much
smaller. Inspection
of Table \ref{tablocal} shows that our local galaxy sample consists of
16 objects in $(U-B)$ and $(U-V)$, and 21 in $(B-V)$.
In the following we indicate with  DS the distant sample of E
galaxies of CL0949 at $z\approx0.4$ and with LS the local sample of 
cluster E galaxies from the literature. 
Figure \ref{figmag} shows the comparison of the cumulative
distributions of the absolute total B band magnitudes of the two
samples, having applied  0.4
mag dimming (cf. Ziegler et al. \cite{ZSBBGS99}) to the DS galaxies.
The distributions look similar 
(the Kolmogorov-Smirnov probability is 0.28), 
although the LS suffers from incompleteness at magnitudes fainter than
$\approx$ -21. On the whole, the  comparison between the two samples 
is fair.

\begin{table*}
\caption[tablocal]{The color gradients of the Local Sample. Column 1
gives the galaxy name, Col. 2 the galaxy cluster (F for Fornax, V
for Virgo, C for Coma), Col. 3 the U-B logarithmic gradient, Col. 4
its error, Col. 5 the U-B color at $R_e/2$, Col. 6 its error, Col. 7
the source, Col. 8 the U-V logarithmic gradient, Col. 9 its error,
Col. 10 the U-V color at $R_e/2$, 
Col. 11 its error, Col. 12 the source, 
Col. 13 the B-V logarithmic gradient, Col. 14 its error, Col. 15 the
B-V color at $R_e/2$, Col. 16 its error,
Col. 17 the source.
The source codes are: F for Franx et al. \cite{FIH89}, G for
Goudfrooij et al. \cite{GHJNDV94}, P for Peletier et
al. \cite{PDIDC90}, I indicates that the V profile was obtained
interpolating the B and R profiles (see text). Errors are due to sky
subtraction and therefore systematic.}
\begin{flushleft}
\begin{tabular}{rrrrrrrrrrrrrrrrr}
\noalign{\smallskip}
\hline
\noalign{\smallskip}
Galaxy   & Cl & a(U-B)&da&U-B&dCol& S & a(U-V) & da & U-V & dCol & S   & a(B-V) & da&B-V &dCol&  S \\
\hline		        				      
  A  496 & - & -0.05 & 0.10 & 0.57 & 0.01 & P & -0.28  & 0.11 &  1.45  & 0.12    & I    & -0.10  & 0.02    & 0.94 & 0.03 & I \\
 IC 1101 & - & -0.05 & 0.10 & 0.69 & 0.01 & P & -0.09  & 0.12 &  1.71  & 0.15    & I    & -0.06  & 0.03    & 0.97 & 0.04 & I \\
NGC 1379 & F & -0.06 & 0.18 & 0.40 & 0.09 & F & -0.02  & 0.28 &  1.35  & 0.03    & I    & -0.05  & 0.04    & 0.92 & 0.02 & I \\
NGC 1399 & F & -0.08 & 0.15 & 0.55 & 0.07 & F & -0.15  & 0.12 &  1.52  & 0.06    & F+G  & -0.07  & 0.01    & 0.97 & 0.02 & G \\
NGC 1404 & F & -0.07 & 0.04 & 0.57 & 0.01 & F & -0.08  & 0.03 &  1.53  & 0.01    & F+G  & -0.01  & 0.01    & 0.96 & 0.01 & G \\
NGC 1427 & F & -     & -    & -    & -    & - &  -     & -    &  -     & -       & -    & -0.05  & 0.01    & 0.89 & 0.01 & G \\
NGC 4365 & V & -     & -    & -    & -    & - &  -     & -    &  -     & -       & -    & -0.04  & 0.01    & 0.97 & 0.01 & G \\
NGC 4374 & V & -0.17 & 0.02 & 0.54 & 0.01 & P & -0.22  & 0.04 &  1.57  & 0.02    & I    & -0.04  & 0.01    & 1.02 & 0.01 & I \\
NGC 4387 & V & -0.01 & 0.02 & 0.49 & 0.01 & P & 0.01   & 0.08 &  1.54  & 0.02    & I    & -0.02  & 0.02    & 1.05 & 0.01 & I \\
NGC 4406 & V & -0.15 & 0.02 & 0.52 & 0.01 & P & -0.16  & 0.03 &  1.52  & 0.02    & I    & -0.02  & 0.01    & 1.00 & 0.01 & I \\
NGC 4472 & V & -0.15 & 0.03 & 0.55 & 0.01 & P & -0.18  & 0.05 &  1.58  & 0.03    & I    & -0.03  & 0.01    & 1.05 & 0.01 & I \\
NGC 4478 & V & -0.12 & 0.04 & 0.52 & 0.01 & P & -0.14  & 0.07 &  1.52  & 0.02    & I    & -0.03  & 0.02    & 0.99 & 0.01 & I \\    
NGC 4486 & V & -0.23 & 0.04 & 0.41 & 0.01 & P & -0.27  & 0.03 &  1.37  & 0.01    & P+G  & -0.04  & 0.01    & 0.96 & 0.01 & G \\
NGC 4564 & V & -     & -    & -    & -    & - & -      & -    &  -     & -       & -    & -0.12  & 0.01    & 0.93 & 0.01 & G \\
NGC 4621 & V & -     & -    & -    & -    & - & -      & -    &  -     & -       & -    & -0.03  & 0.01    & 0.92 & 0.01 & G \\
NGC 4636 & V & -0.22 & 0.04 & 0.46 & 0.01 & P & -0.27  & 0.07 &  1.45  & 0.05    & I    & -0.05  & 0.02    & 0.99 & 0.01 & I  \\
NGC 4649 & V & -0.15 & 0.04 & 0.65 & 0.01 & P & -0.18  & 0.06 &  1.69  & 0.03    & I    & -0.04  & 0.01    & 1.04 & 0.01 & I  \\
NGC 4660 & V & -     & -    & -    & -    & - &  -     & -    &  -     & -       & -    & -0.05  & 0.01    & 0.98 & 0.01 & G  \\
NGC 4874 & C & -0.07 & 0.06 & 0.55 & 0.01 & P & -0.20  & 0.05 &  1.50  & 0.05    & I    & -0.10  & 0.02    & 0.97 & 0.01 & I  \\
NGC 4889 & C & -0.08 & 0.04 & 0.72 & 0.01 & P & -0.11  & 0.09 &  1.65  & 0.04    & I    & -0.10  & 0.03    & 0.95 & 0.02 & I  \\
NGC 6086 & - & -0.15 & 0.04 & 0.62 & 0.01 & P & -0.17  & 0.14 &  1.58  & 0.06    & I    & -0.03  & 0.03    & 0.96 & 0.01 & I  \\
\hline
\end{tabular}
\end{flushleft}
\label{tablocal}
\end{table*}

\begin{figure}[ht!]
\resizebox{\hsize}{!}{\includegraphics{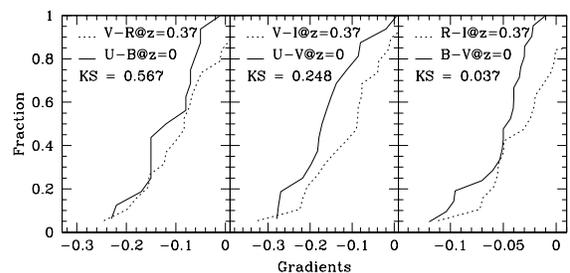}}
\caption[Cumulative distributions of color gradients]
{Cumulative distributions of color gradients of the z=0 
(solid) and $z\approx 0.4$ (dotted) samples. The left, central and
right panels show respectively the $(U-B)$, $(U-V)$ and $(B-V)$
gradients of the LS, to be compared to the $(V-R)$, $(V-I)$ and
$(R-I)$ gradients of the DS. Each panel is labeled with the results 
of the KS tests applied to the corresponding pairs of distributions.}
\label{figcompgrad}
\end{figure}

Fig. \ref{figcompgrad} compares the cumulative distributions of the
measured gradients for the LS and the DS samples. 
The distributions appear rather similar, suggesting little
evolution. In particular, the median values of the $(U-B)$, $(U-V)$
and $(B-V)$ gradients of the LS galaxies are -0.12, -0.17 and -0.06
mag/dex per decade respectively. These are close to, but somewhat steeper
than the median
values of the $(V-R)$, $(V-I)$, and $(R-I)$ gradients of the DS
galaxies, that are respectively -0.09, -0.08, -0.03 mag/dex
per decade.  A Kolmogorov-Smirnov
(KS) test shows that only the (B-V), (R-I) pair of distribution is
marginally different.

In the following we will investigate quantitatively the evolutions of
color gradients under different assumptions for their origin and
compare the models to the observed distributions.

\section{The Models}
\label{modeling}
\subsection{The Basic Tool: Simple Stellar Populations}
\label{ssps}

Our modeling is based on the assumption that the
light-averaged stellar population in the Es between $z\approx 0.4$ and
$z=0$ can be described by Simple Stellar Populations (SSP), i.e. coeval
and chemically homogeneous assemblies of single stars.  The SSP models
for this work are computed with the evolutionary synthesis code
described in Maraston (\cite{Ma98}), in which the Fuel Consumption
theorem (Renzini \& Buzzoni \cite{RB86}) is used to evaluate the
energetics of Post Main
Sequence stars. The input stellar tracks for the SSP models used
here are from Cassisi ({\it private communication}; see also
Bono et al. \cite{bono97}), covering the metallicity range ${\rm
[Fe/H]}=[-1.35\div0.35]$, with helium-enrichment parameter 
${\rm\Delta Y/\Delta Z}$=2.5. The amount of mass loss along the Red Giant
branch (RGB) is parametrized a' la Reimers (\cite{R75}), with the efficiency
$\eta$=$0.33$ as calibrated by Fusi Pecci \& Renzini (\cite{FPR76}). 
The mass loss prescriptions affect the
evolutionary mass on the horizontal branch (HB), leading to warmer HBs
as the mass loss increases.
We adopt Salpeter IMF ($\Psi(m)=A{m}^{-(1+x)}$, $x$=1.35) down to
a lower mass limit of 0.1 M$_\odot$. 
For plausible variation of the IMF slope, the optical
colors of SSPs are virtually unchanged (e.g. Maraston \cite{Ma98}).  
The evolutionary synthesis code has been updated for the computation
of the SSP Spectral Energy Distributions (SEDs), as a function of $t$ 
and [Fe/H]. We  adopted the spectral library of Lejeune, Cuisinier and 
Buser (\cite{LCB98}) to describe the stellar spectra as
functions of gravity, temperature and metallicity.
A redshifted grid of SSP model SEDs has also been computed to
interpret the colors of the DS galaxies. We have applied the same
redshift $z=0.3775$ to all the models. This is the average value of
the $z$ distribution of the DS sample, having assigned to the galaxies
without redshift determination, the average redshift value of the
objects with a measured redshift. The uncertainties introduced by a
$\approx 0.03$ variation in redshift are of the same order as the
statistical errors.
Model SEDs have been convolved with the $UBVRI$ Johnson-Cousins filter
functions (Buser \cite{B78}, Bessel \cite{B79}) to yield synthetic 
magnitudes. The applied zero points are
($U$,$B$,$V$,$R$,$I$)=(0.02, 0.02, 0.03, 0.0039, 0.0035)  
for the Vega model atmosphere of Kurucz (\cite{K79}) with 
($T_e$,~g,~[Fe/H])=(9400,~3.9,~-0.5). This
ensures consistency with our HST photometry (Holtzman et al. 1995).
We refer to Maraston (\cite{Ma99}) for more details.  

\begin{figure*}[ht!]
\resizebox{\hsize}{!}{\includegraphics{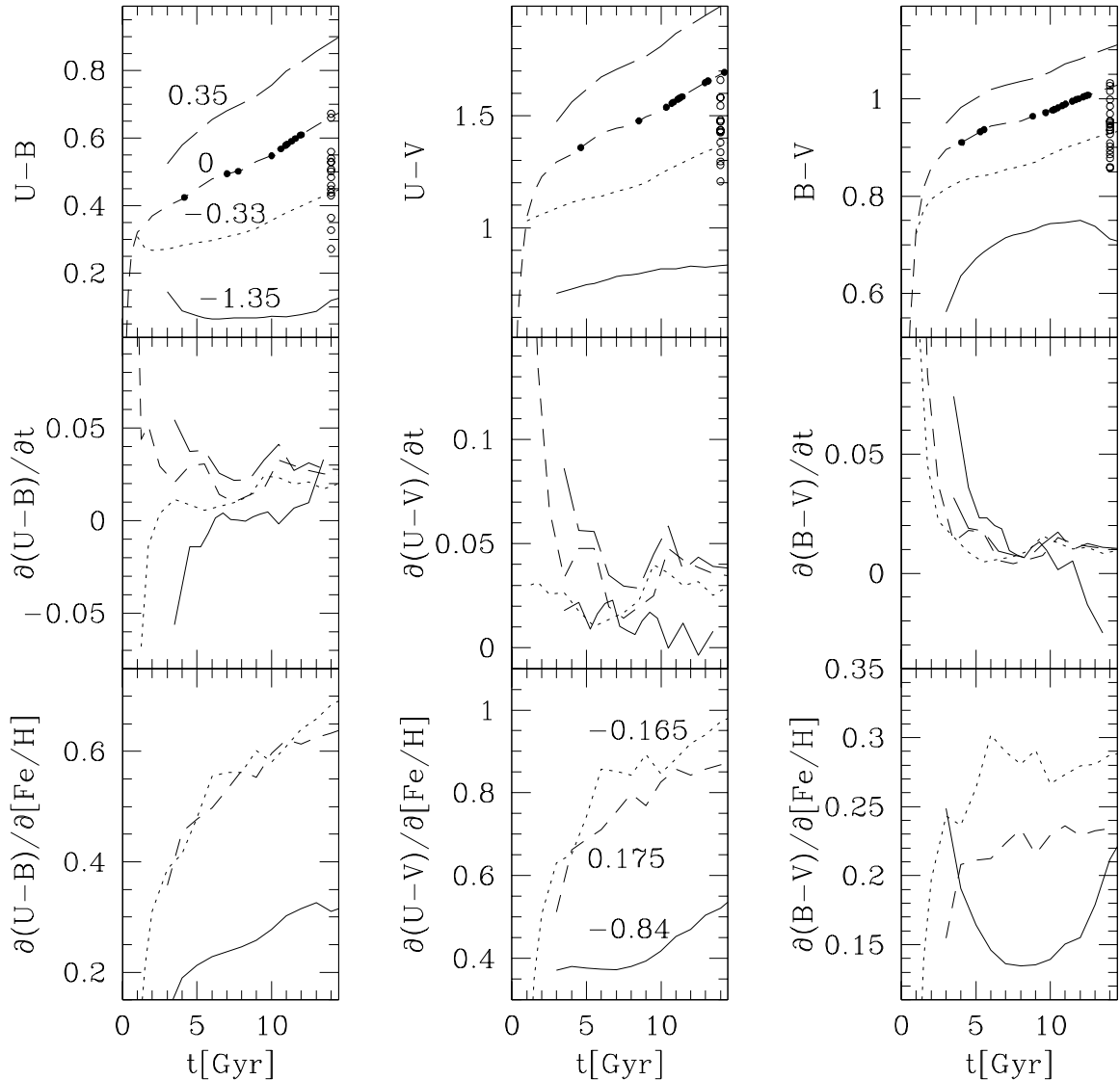}}
\caption[Models of color gradients]{The colors, time and [Fe/H]
derivatives of the rest-frame SSP models in the U, B and V bands. The
different line type of the color and time derivative rows 
refer to the metallicities indicated in the top-left plot. The line
types
and their corresponding metallicities (midpoints of the above)
of the [Fe/H] derivative are given in the bottom panel of the
central figure. The filled dots show the 0.2 $R_e$ colors of the LS
galaxies with ages assigned on the solar metallicity line. The open
dots show their 2 $R_e$ colors at the (arbitrary) age of 14 Gyr.}
\label{figmodelz0}
\end{figure*}

\begin{figure*}[ht!]
\resizebox{\hsize}{!}{\includegraphics{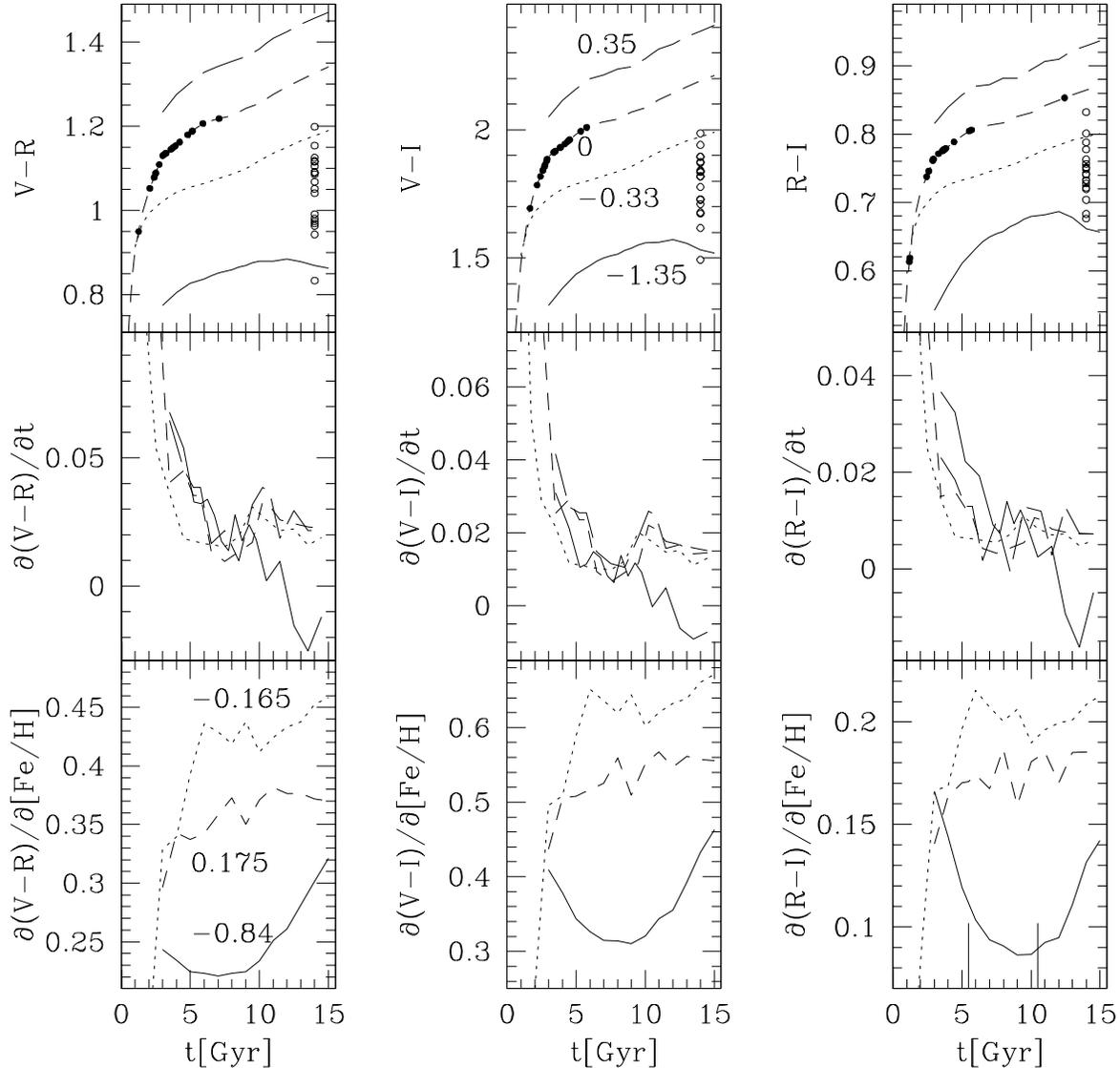}}
\caption[Models of color gradients]{The colors, time and [Fe/H]
derivatives of the redshifted SSP models in the $V,~R,~I$ bands. The
different line type of the color and time derivative rows 
refer to the metallicities indicated in the top-middle plot. 
The line types 
and their corresponding metallicities (midpoints of the above)
of the [Fe/H] derivative are given in the bottom-left plot.
The vertical lines show the age (5.47 and 10.47 Gyr) 
of galaxies at z=0.3775 that today are 10 and 15 Gyr old,
respectively. The filled dots show the 0.2 $R_e$ colors of DS
galaxies with ages assigned on the solar metallicity line. The open
dots show their 2 $R_e$ colors at the (arbitrary) age of 14 Gyr.}
\label{figmodelz04}
\end{figure*}

Figures \ref{figmodelz0} and \ref{figmodelz04} summarize the
properties of the models relevant to our case. 
Fig. \ref{figmodelz0} shows the $(U-B)$, $(U-V)$ and $(B-V)$ colors of
the rest-frame SSP models, together with   
their derivatives with respect to age and metallicity (given as
[Fe/H]). Derivatives are computed as finite differences on the grid
and assigned to the midpoints in age and metallicity.

In general, optical colors become redder with increasing age 
and/or increasing metallicity. 
In our models, there are two exceptions to this trend, both at low
metallicities. The [Fe/H] = -1.35 models become bluer in $U-B$ as they
age from $\sim$ 3 to $\sim$ 5 Gyr. This is due to 
the properties of the $U-B$ stellar colors for $T_{\rm e} \sim $ 8500-7500 K
(see e.g. Castelli \cite{C99}) typical of turnoff stars 
in this age range. 
At higher metallicities this effect does not appear, because when the
turnoff stars have temperatures in this range, their contribution to
the total light is less important due to the presence of AGB stars
(Maraston \cite{Ma99}). 

At old ages, the same set of models become bluer and bluer in $B-V$
from $\sim$ 12 Gyr onward, reflecting warmer and warmer HBs. 
With $\eta=0.33$, this effect sets in for [Fe/H]$\la -0.5$. 
It disappears when no mass loss
is applied (Maraston \cite{Ma99}; see also Chiosi et al. \cite{CBB88}).
The time derivative of colors shows a noisy behavior due to the
discrete modeling. However, a general trend can be appreciated, with
the derivative initially decreasing and leveling off at $\sim$ 0.01
-- 0.04 mag/Gyr for ages older than $\sim$ 7 Gyr. The time derivatives
appear almost independent on metallicity for $Z \ge$ 0.5
Z$_\odot$, while the lowest $Z$ set of models behave differently due
to the two effects described above.

The derivatives of colors with respect to metallicity show a different
trend with age, with colors generally becoming more and more sensitive to
[Fe/H] as the SSPs get older. It is worth noticing that the plotted
curves are color variations per metallicity decade. The dependence on
$Z$ is quite different, with low metallicity model colors much more
sensitive to $Z$ variations than the high $Z$ ones. We preferred to
plot the derivatives with respect to [Fe/H] because these are the
actual tools used in our simulations. We also point out that in our
models the helium abundance varies in lock-steps with $Z$. This
influences the temporal behavior of the SSP colors and of their
derivatives, especially at high metallicity.

\begin{figure*}[ht!]
\resizebox{\hsize}{!}{\includegraphics{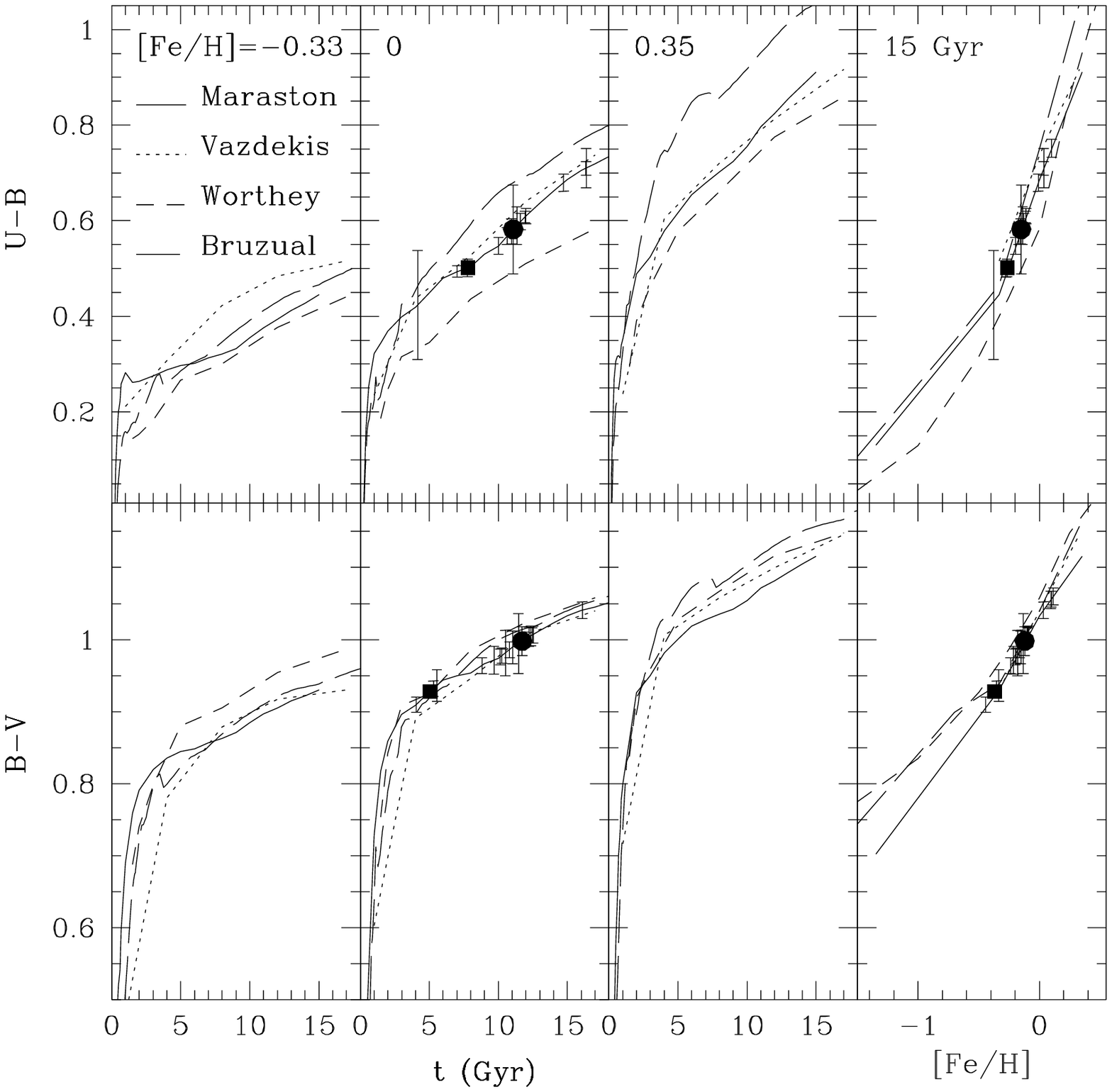}}
\caption[Comparison of models]
{The $(U-B)$ (top) and $(B-V)$ rest-frame color time evolution of our
model (full lines), Vazdekis et al. (\cite{Vetal96}, dotted lines),
Worthey (\cite{W94}, short-dashed lines), Bruzual \& Charlot (\cite{BC99},
long-dashed lines) as a function of age (for three metallicities) 
and metallicity (15 Gyr age for our models and Bruzual \& Charlot, 17
Gyr for Vazdekis and Worthey's models). The error bars show the
allowed range of the inner (at $0.2 R_e$) colors of the LS galaxies 
derived from the gradient fits. 
The filled circles and squares show the inner and outer ($2R_e$)
colors, respectively, of the galaxy NGC 1399.}
\label{figcompmod}
\end{figure*}

Figure \ref{figcompmod} compares our models (shown down to younger ages
for the purpose) to the ones of 
Vazdekis et al. (\cite{Vetal96}), Bruzual \&
Charlot (\cite{BC99}) and  Worthey (\cite{W94}) as a function of
metallicity and age (see Maraston \cite{Ma98} for a comparison to the
models of Tantalo et al. \cite{Tetal96} at solar metallicity). 
On the whole, the agreement between all the models in the $(B-V)$
color is good. The variations between the models are of the same order of
the uncertainties in the observed colors. 
Our SSP optical colors 
compare remarkably well with Vazdekis et al. (\cite{Vetal96}).
Larger differences between the models are observed in the $(U-B)$.
Bruzual \& Charlot (\cite{BC99}) models tend to have systematically
redder $(U-B)$.
Worthey (\cite{W94}) models produce instead systematically bluer
$(U-B)$ (and $(U-V)$) colors at a given $(B-V)$. A thorough
investigation of these discrepancies will be presented in Maraston 
(\cite{Ma99}). The 
age (at fixed metallicity) or metallicity (at fixed age) differences
inferred using the  different models between
the central and outer parts of the galaxies (the galaxy NGC 1399 is
shown as an example) are consistent within the errors.
The results on the origin of the color gradients presented here also
depend on the derivatives of the colors with respect to age and
metallicity. This dependence is difficult to estimate quantitatively
because of the coarseness of the grid of the other sets of models, and
of the unavailability of their entire SEDs. The close
similarity of our models to the Vazdekis is reassuring.

Similar to Fig. \ref{figmodelz0}, Fig. \ref{figmodelz04} shows the 
properties of the redshifted SSP
SEDs in the $V$, $R$ and $I$ bands. Broadly speaking, the $(V-R)$,
$(V-I)$ and $(R-I)$ colors trends with age resemble those of the $(U-B)$,
$(U-V)$ and $(B-V)$ colors. Some differences are present, as a result
of the different effective wavelengths and sensitivity curves of the
corresponding pairs of filters. In particular we notice that the time
derivative of the $(V-I)$ color is approximately a factor of 2 smaller
than that of the $(U-V)$ color, reminiscent of Fig. \ref{figcompgrad}. 
Thus we expect that the $(U-V)$ colors
of nearby galaxies are more sensitive to age gradients than the
$(V-I)$ colors of distant objects.

Simple inspection of Figs. \ref{figmodelz0} and
\ref{figmodelz04} already allows us to define the framework for the
interpretation of color gradients in ellipticals. 
The filled and open dots show the 0.2 and 2 $R_e$ colors
of the LS and DS galaxies. It is clear that the range of ages and 
metallicities 
considered here is adequate to model the observed colors both as a
pure age sequence (at a fixed metallicity) or as a pure metallicity
sequence (at a fixed age).

Turning now to the gradients' evolution, we start considering that the
LS galaxies all (with one exception) 
have negative gradients, with colors being redder in
the central regions. Thus, if they are caused by {\it
age gradients only}, centers must be {\it older} than the 
outer parts ($\approx 5$ Gyr per radial decade). In addition, 
if the present age of the Es
centers is large enough ($\approx 15$ Gyr), we expect the age
derivatives of the colors to stay approximately constant for look-back
times up to $\approx 7$ Gyr, and therefore to measure color gradients
in the DS of approximately the same size in $(V-R)$ and $(R-I)$, and 
twice as small in $(V-I)$. 

In contrast, if color gradients are caused by 
{\it metallicity gradients only}, the centers
of local Es must be more metal rich than the outer parts ($\approx 1.6$
more metal rich per radial decade).  In addition, given the increase
with age of the metallicity derivatives of the colors, we expect the
color gradients of the DS to be flatter than those of the LS. 
This effect should be less prominent if the centers of ellipticals are old. 

As noted in Sect. \ref{cl0952}, 
some of the gradients measured in CL0949 Es are positive. In
Sect. \ref{errors} we shall argue that these are caused by measurement
errors, with an underlying distribution of intrinsically negative
gradients. Here we note that a change in sign with time of the color
time derivative of SSP models is possible only at low
metallicities and/or low ages. 
Color gradients caused by metallicity 
cannot change sign with time following passive evolution. 

\subsection{Modeling the Evolution of Color Gradients}
\label{cgmodels}

The detailed modeling of the color gradients evolution of each 
galaxy involves a number of steps. As a starting point, a 
cosmological model should be
chosen to fix the look-back time at $z\approx 0.4$.  Here we adopt
${\rm H_{o}}$=65 ${\rm Km\cdot s^{-1}\cdot Mpc^{-1}}$ and $\rm
{\Omega_{m}}$=0.2, with ${\rm {\Lambda}}=0.8$. The age of the
Universe is $\approx$ 16.18 {\rm Gyr} and $t_{\rm lb}=4.53$
{\rm Gyr} at $z$=0.3775.  The choice of the cosmological parameters is
not crucial to model color gradients, but a much 
younger universe (for example, 10 Gyr) would strongly disfavor the 
age hypothesis for the origin of color gradients (see below).
In addition, the look-back time
has an influence on the exact value of the predicted colors. As we
will show later, the
optimal agreement between the observed color distributions in the LS
and the DS samples is achieved for larger look-back times, i.e.,
$t_{\rm lb}\approx 8$ Gyr. 

Inspection of Figs. \ref{figgrad555675}-\ref{figgrad675814} shows that
the color gradients of the DS galaxies are measured in the radial
range $0.2-2~R_e$.  The same applies to most of the local
galaxies. Therefore, we choose this radial decade to model the
gradients. A more innerly defined decade (i.e., $0.1-1~R_e$) suffers
of insufficient resolution for the more distant sample. A more outerly
defined decade (i.e., $0.3-3~R_e$) is poorly investigated in the local
sample. Thus, we analyse a given color gradient by examining
the colors at the $0.2~R_e$ (hereafter, the central region) and
$2~R_e$ (hereafter, the outer region) isophotes, as derived from the
values listed in the Table \ref{tabresultsvr}-\ref{tablocal}.

We need now to choose representative SSP models for the inner and
outer regions. Due to the age-metallicity degeneracy, one parameter
has to be fixed to derive the other from the observed colors. For the
central stellar populations we consider two options: 1) Case m, in which
all the galaxy centers have the same age and the colors are used to
infer their metallicities; 2) Case a, in which all the galaxy centers
have the same metallicity, and their colors are used to infer their
ages. In Case m we explore two possible current ages: 10 and 15 Gyr;
in Case a we assign solar metallicity to all the galaxy centers. A
smaller metallicity (i.e., half-solar) would imply implausibly large
(larger than the age of the Universe) ages for the galaxies (see
Figs. \ref{figmodelz0} and \ref{figmodelz04}). The
canonical interpretation of the Color-Magnitude relation of cluster
early type galaxies is consistent with Case m 
(e.g. Arimoto \& Yoshii \cite{AY86}, 
Kodama \& Arimoto \cite{KA97}); Case a is nevertheless worth
exploring.

As for the gradients, we consider the two extreme possibilities that
they are driven by pure age ({\bf A}) or pure metallicity ({\bf M}) 
gradients. Thus we have four scenarios in which we model the evolution
of the color gradients:

\begin{itemize}
\item[a{\bf A}:] age drives the color gradients. Central and outer
SSPs have solar $Z$ and colors
reflect their ages. The gradients are expected to steepen with
look-back time if the centers are young; to remain nearly constant
if they are old enough. 
\item[m{\bf A}:] age drives the color gradients. The central SSPs of
local Es are 10 or 15 Gyr old (m{\bf A}10 and m{\bf A}15);
the outer SSPs have the same metallicity as the central ones, but
different ages. The gradients' evolution is expected to be similar to case
a{\bf A}.
\item[a{\bf M}:] metallicity drives the color gradients. 
Central SSPs have solar $Z$; the outer SSPs have
the same age as the centers, but different metallicities. Color
gradients are expected to flatten with look-back time.
\item[m{\bf M}:] metallicity drives the color gradients. 
The central and outer SSPs of local Es are coeval and 
10 or 15 Gyr old (m{\bf M}10 and m{\bf M}15), 
and the colors reflect their metallicities.
The gradients' evolution is expected to be similar to case a{\bf M}.
\end{itemize}

To summarize, we model the evolution of the color gradients of each
galaxy in the LS in the four scenarios by: 

\begin{itemize}
\item[1.)] fixing the age and metallicity of the central and outer SSPs
according to the specific option (a{\bf A},a{\bf M},m{\bf A} and
m{\bf M});
\item[2.)] rejuvenating the SSPs by the adopted look-back time (4.53 Gyr);
\item[3.)] computing the color gradients of the rejuvenated and redshifted SSPs
in the $V,~R$ and $I$ bands.
\end{itemize}

For consistency, we also perform the reverse experiment and compute
the predicted $U,B,V$ color gradients of the CL0949 galaxies after
$t_{LB}$ (i.e., 4.53 Gyr) of passive evolution.
We tested that the results we are going to discuss remain the same 
within the errors if the
whole color profiles are modeled at each radial distance, and a
logarithmic gradient is fit to the predicted profiles at z=0.

\section{Ages and Metallicities of the Galaxies in the Four
Scenarios}
\label{agesandzs}
 
We consider now the applicability of the four scenarios to our local
and distant galaxy samples. Indeed, 
the ages or metallicities derived from the three colors
available should agree within the measurement errors. The distribution
of ages derived for cases a{\bf A}, a{\bf M} and m{\bf A} 
for the LS and the DS galaxies
should be similar, given the look back time. The distributions of
metallicities derived for cases a{\bf M}, m{\bf A} and m{\bf M} 
should also be similar.

\subsection{Age gradients: the a{\bf A} Case}
\label{AA}

\begin{figure}[ht!]
\resizebox{\hsize}{!}{\includegraphics{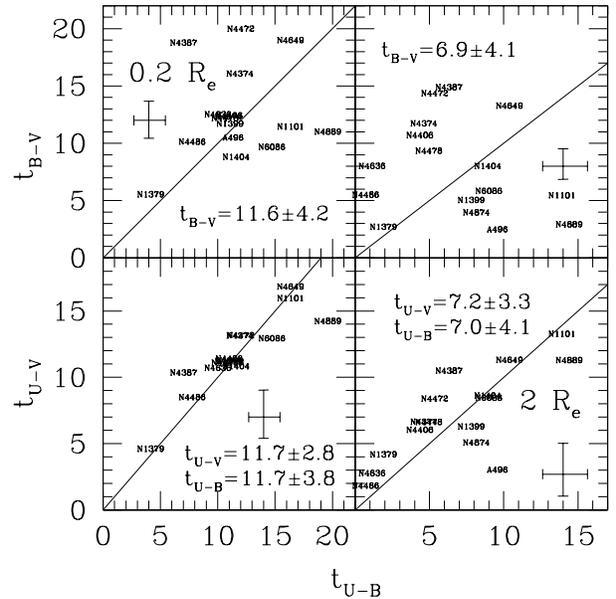}}
\caption[Color gradients]{Ages assigned to the center (0.2 $R_e$) 
and the outer isophotes (2 $R_e$) of the local galaxies, assuming that they 
have solar metallicity. The given mean age and rms are derived using
all the galaxies available in a given filter. The cross shows the mean
of the measurement errors.}
\label{figlocaa}
\end{figure}

\begin{figure}[ht!]
\resizebox{\hsize}{!}{\includegraphics{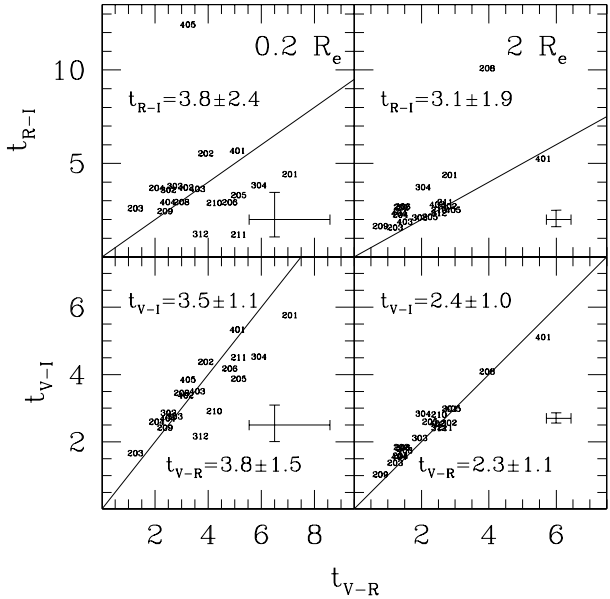}}
\caption[Color gradients]{Ages assigned to the center (0.2 $R_e$) 
and the outer isophotes (2 $R_e$) of the distant galaxies, assuming that they 
have solar metallicity. Labels and crosses as in Fig. \ref{figlocaa}}
\label{fighigaa}
\end{figure}

Figures \ref{figlocaa} and \ref{fighigaa} show the ages assigned to
the center (0.2 $R_e$) and the outer isophotes (2 $R_e$) of the local
and distant galaxies, having assumed solar metallicity for both
samples in both regions. The ages derived from different colors are in 
agreement (within the rather large statistical errors), better for the
DS objects. 
We note that the LS objects appear in the plot only when the
specific pair of colors is available. However, the average ages and
rms indicated in the plots are those obtained considering all the
objects with the specified color available. 
On the average, the centers of LS galaxies appear $\sim$ 12 Gyr old,
while the peripheries are $\sim$ 7 Gyr old. The DS galaxies appear
$\sim$ 4 Gyr old in the center, $\sim$ 2.5 Gyr old in the
peripheries. Thus, the consistency of the a{\bf A} hypothesis is
rather questionable, both because the average age difference between the
centers of the two samples is larger than our adopted look-back time,
and because the age gradients implied under this scenario are much
smaller for the DS sample than for the LS one.
In addition,  the colors of
some local galaxies require a present age
younger than 4.5 Gyr or an age older
than the age of the universe in our chosen cosmology.
In other words, only a subsample of the LS galaxies could be the
passively evolved descendants of the CL0949 objects. This is partly a
consequence of forcing a solar $Z$ for all the considered SSPs.

\subsection{Metallicity gradients: the m{\bf M} Case}
\label{MM}

\begin{figure}[ht!]
\resizebox{\hsize}{!}{\includegraphics{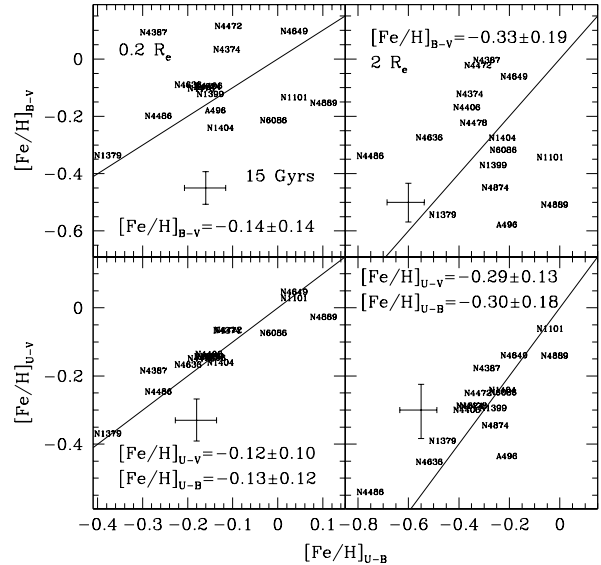}}
\caption[Color gradients]{Metallicities assigned to the center (0.2 $R_e$) 
and the outer isophotes (2 $R_e$) of the local galaxies, assuming that
their age is 15 Gyr. The given mean metallicity and rms are derived using
all the galaxies available in a given filter. The cross shows the mean
of the measurement errors.}
\label{figlocmm}
\end{figure}

\begin{figure}[ht!]
\resizebox{\hsize}{!}{\includegraphics{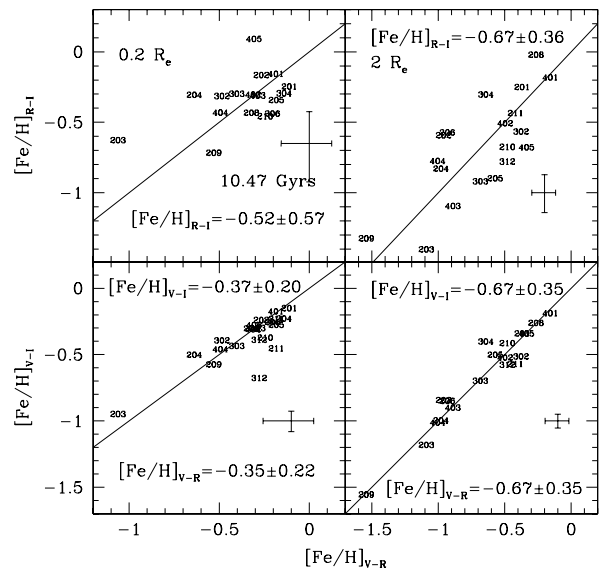}}
\caption[Color gradients]{Metallicities assigned to the center (0.2 $R_e$) 
and the outer isophotes (2 $R_e$) of the distant galaxies, assuming
that their age is 10.47 Gyr. Labels and crosses as in Fig. \ref{figlocmm}.}
\label{fighigmm}
\end{figure}

Figures \ref{figlocmm} and \ref{fighigmm} show the metallicities
assigned to the center (0.2 $R_e$) and the outer isophotes (2 $R_e$)
of the local and distant galaxies, assuming a present day
age of 15 Gyr for the LS objects.  
At $z=0$ the metallicities derived from the three
colors agree well.
Local galaxies appear to be slightly subsolar at their
centers if 15 Gyr old, solar if 10 Gyr old.
The central SSPs of the DS objects require only half solar
metallicity, if the look-back time is 4.53 Gyr, but with t$_{\rm lb}
\approx$ 8 Gyr the central $Z$ is larger by 0.15 dex.
This discrepancy is the same
present in the a{\bf A} case: for our adopted look-back time the
observed central colors of the DS objects are too blue (those of the LS are
too red) when compared to passively evolving models.
The metallicity gradients are $\approx$ 0.2 dex for both DS and LS, 
indicating that metallicity
driven color gradients are a viable interpretation (see Sect. \ref{results}).
However, we notice that the DS galaxies span a wider range in
metallicity gradients, compared to LS objects. This indicates a slight
inconsistency also for the m{\bf M} scenario, but see Sect. \ref
{errors} for a possible way out.

\subsection{The a{\bf M} and m{\bf A} Cases}
\label{AMandMA}

The a{\bf M} case is always applicable:
since the central ages derived in \ref{AA} are $\approx 12$
Gyr on the average, and the central metallicities are supposed to be
solar, the other colors for both DS and LS galaxies can be
reproduced consistently in all filters by decreasing the metallicity
by 0.2 dex. 
 
The applicability of the m{\bf A} scenario strongly depends on 
the assumed central age. When 10 Gyr is adopted for the centers of
LS galaxies, a substantial number of objects appear to have outer SSPs
younger than our adopted look-back time.
In addition, the average age of the outer regions of the DS galaxies
is $\approx$ 3 Gyr, too old to evolve into the LS galaxies. 
The situation is different if the central age of local galaxies is
fixed to 15 Gyr. 
The average outer age of the LS is $\approx$ 10 Gyr in this
case, with only a few objects younger than our adopted look-back
time. The peripheries of the DS galaxies appear $\approx$ 6 Gyr old on
the average, which is consistent with the average age of the LS. 

To summarize, the metallicity driven color gradients hypothesis is
always applicable, while the age hypothesis holds only if local galaxies
are old ($\sim$ 15 Gyr). This reflects the fact that a small
metallicity variation is required to explain the observed color gradients,
while a large  age gradient (of $\sim$ 5 Gyr) is needed to fit the
color gradients of the LS objects. With a 4.5 Gyr look-back time, a
substantial number of LS galaxies do not have their corresponding
object in the DS.   

\section{Results}
\label{results}

In this section we present the results of the detailed modeling of the
evolution of the central colors and color gradients under the four
scenarios, according to the procedure outlined in
Sect. \ref{cgmodels}.
We will compare the model predictions to the data by considering the
cumulative distribution functions of the color gradients and applying
Kolmogorov-Smirnov tests to the pairs of correspondent theoretical and
observational distributions. In particular, the redshifted $V,~R,~I$
colors of rejuvenated LS galaxies will be compared to those of CL0949
objects; the $U,~B,~V$ colors of passively evolved DS galaxies will be
compared to the $U,~B,~V$ colors of local ellipticals. As
discussed above, the colors of some LS galaxies are so blue that ages
shorter than $t_{LB}$
would be required. In these cases the object ``drops out'' of the
sample by getting an assigned (blue) color of -1. For some
choices of age or metallicity the colors are outside our grid of
models. In these cases we assign the minimum (-1.35) or maximum (0.35)
metallicities, and 
30 Myr of age (the smallest age of the grid), 
if the extrapolated age is smaller
than 0, or 15 Gyr, if larger ages are extrapolated.

\subsection{The distributions of central colors}
\label{distcol}

\begin{figure}[ht!]
\resizebox{\hsize}{!}{\includegraphics{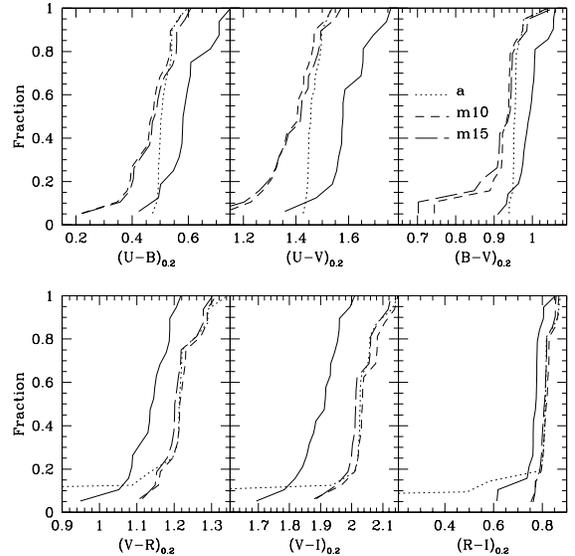}}
\caption[Color gradients]{The cumulative distributions of observed (solid line)
and simulated (for $t_{LB}=4.53$ Gyr) 
central colors of z=0 (top panels) and $z\approx 0.4$
(bottom panels) galaxies. The dotted lines show Case a, the short- and 
long-dashed line the Case m for present-day central ages of 10 and 15
Gyr. The Kolmogorov-Smirnov probabilities that the
observed and simulated data are drawn from the same
underlying distribution are always smaller or equal than 0.001.}
\label{figdistcol}
\end{figure}

Fig. \ref{figdistcol} compares the cumulative distributions of
observed and simulated central colors of local and distant galaxies.
The Kolmogorov-Smirnov probability {\rm KS}
that the observed and
simulated data are drawn from the same underlying distribution are
always smaller or equal than 0.001. None
of the simulated central color distributions appear compatible with
the observed ones. This result is confirmed using the Wilcoxon
test. The simulated centers of the DS objects appear too
blue (by $\approx 0.04$ mag in $B-V$) when compared to the local ones,
and, vice versa, the centers of local galaxies appear too red 
($\approx 0.03$ mag in R-I) when compared to the CL0949 galaxies. The
differences are more pronounced in the $(U-B)$ and $(V-R)$ and $(U-V)$ and
$(V-I)$ pairs. Measurement errors alone seem insufficient to explain the
difference between these pairs (see \ref{errors}), while the $(B-V)$
and $(R-I)$ colors might not differ significantly in a statistically sense.
The uncertainties in the galaxy redshifts and in the calibration to
the Johnson-Cousins system (see Holtzmann et al. \cite{HBCHTWW95}), 
combined with
the notoriously difficult photometric calibration of the U band
probably suffice to explain our findings. In any case, we tested 
that the conclusions concerning the
origin of color gradients are not affected by this possible inconsistency
of the central colors. The distributions of gradients shown in
Fig. \ref{figdistgrad} are hardly changed when the look-back time is
increased from 4.53 to 8 Gyr.

Assuming that the color differences are real,
they can be explained using a larger
look-back time. For example, $t_{LB}=8$ Gyr (obtained with the same
cosmology and a rather extreme low value of the Hubble constant, $H_0=37$) 
gives reasonable KS probabilities (5-80\% for the different colors and models).
The result might also point to
recent, minor episodes of star formation in the distant galaxies
(Poggianti et al. \cite{Pog98}, Couch et al. \cite{Cou98}). 

\subsection{The distributions of color gradients}
\label{distgrad}

\begin{figure}[ht!]
\resizebox{\hsize}{!}{\includegraphics{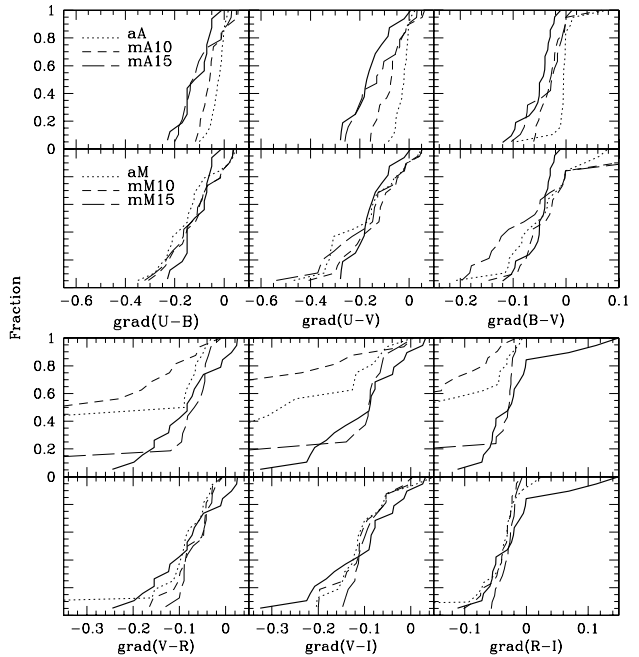}}
\caption[Color gradients]{The cumulative distributions of observed
and simulated color gradients of z=0 (top panels) and $z\approx 0.4$
(bottom panels) galaxies. In each of the plots the thick solid line shows
the observed distributions. Two rows of plots are shown for the local
and distant case. The top row refers to the age-driven color
gradients. Here the dotted lines show the a{\bf A} case, the
short-dashed lines the m{\bf A}10 case, the long-dashed lines the m{\bf
A}15 case. The bottom row refers to the metallicity-driven color
gradients. Here the dotted lines show the a{\bf M} case, the
short-dashed lines the m{\bf M}10 case, the long-dashed lines the m{\bf
M}15 case. The Kolmogorov-Smirnov probabilities that the
observed and simulated data are drawn from the same underlying
distribution are given in Table \ref{tabKS}.}
\label{figdistgrad}
\end{figure}

Fig. \ref{figdistgrad} compares the cumulative distributions
of the observed and 
simulated color gradients.
The Kolmogorov-Smirnov probabilities that the
observed and simulated data are drawn from the same
underlying distribution are given in Table \ref{tabKS}. 

The a{\bf A} Case is clearly ruled out. The very small differences between
the central and outer ages required at $z\approx 0.4$ (see
Fig. \ref{fighigaa}) evolve into too shallow gradients at z=0. Along
the same line, the 5 Gyr age difference inferred at z=0 (see
Fig. \ref{figlocaa}) produces too steep gradients at $z\approx 0.4$,
because the color time derivatives increase for decreasing age (see
Fig. \ref{figmodelz0} and discussion in Sect. \ref{ssps}).

The m{\bf M} case is viable for both the assumed present-day central ages of
10 and 15 Gyr. The t-test confirms that the means of the observed and
simulated distributions in all colors at both z=0 and $z\approx 0.4$
are statistically identical. The f-test on the rms of the
distributions confirms the slightly lower KS probabilities obtained 
for the  $B-V$ gradients for z=0 and for the
$(R-I)$ for z$\approx 0.4$. 
As discussed in Sect. \ref{ssps}, all (except one) local gradients
are negative, while some positive gradients are observed at $z\approx
0.4$. This makes the observed widths of the $z\approx 0.4$ gradient 
distributions
larger than the simulated ones. Errors might partly explain this effect
(see discussion in Sect. \ref{errors}). Indeed, the widths of the
distributions are somewhat dissimilar, with the DS presenting a tail
of steep gradients with no counterpart in the simulated LS. Since the 
metallicity
color derivatives decrease with decreasing age, the evolution leads to
shallower (not steeper) gradients. The same considerations apply
to the a{\bf M} Case, which provides also an acceptable model of color
gradients. 

The m{\bf A} Case does not offer a perfect 
clear-cut solution. One can clearly exclude the case of central
present-day young ages (10 Gyr), since the resulting outer ages at z=0
are either younger than the assumed $t_{LB}$, or so young to
correspond to a very steep rejuvenated color gradient. 
The outer colors of the distant galaxies require such large ages that
too shallow gradients
are produced at z=0. In contrast, present day central ages
of 15 Gyr still make the age hypothesis a possible explanation
for color gradients. In this case, the outer ages (some $\approx 5$
Gyr smaller than the central ones) required to match
the colors of the z=0 galaxies are large enough (see Sect. \ref{AMandMA}) 
to produce mild gradients at $z\approx 0.4$, even if a tail of too
steep  gradients is produced, which is not observed in the
data. Similarly, only a small fraction ($\approx 10-20$\%) of the
distant galaxies cannot be matched in their outer colors by this
mechanism. The remaining objects are assigned outer ages that produce
reasonable gradients locally.

\begin{table*}
\caption[KS]{The Kolmogorov-Smirnov probabilities of the color
gradient distributions. Col. 1 gives the examined Case, Cols.2-5 the KS
probabilities for the $(U-B)$, $(U-V)$, $(B-V)$ color gradients at
$z=0$, Cols. 6-7 for the $(V-R)$, $(V-I)$, $(R-I)$ at $z\approx0.4$.}
\begin{flushleft}
\begin{tabular}{rrrrrrr}
\noalign{\smallskip}
\hline
      Case         &KS$_{U-B}$ &  KS$_{U-V}$ &  KS$_{B-V}$  & KS$_{V-R}$&KS$_{V-I}$ & KS$_{R-I}$\\
\hline
     a{\bf A}  & 7.5e-05 & 4.4e-06 & 2.9e-07 &     0.05 &  0.013 &  0.002\\
   m{\bf M}10  &    0.79 &   0.50 &    0.18 &	  0.62 &   0.50 &     0.18\\
   m{\bf M}15  &    0.67 &   0.51 &    0.09 &	  0.17 &   0.29 &     0.22\\
     a{\bf M}  &    0.37 &   0.15 &    0.20 &	  0.83 &   0.50 &     0.32\\
   m{\bf A}10  &   0.016 &   0.004 &    0.08 &	 0.009 &  0.0002 &   2.7e-05\\
   m{\bf A}15  &    0.83 &   0.50 &    0.01 &	  0.62 &   0.87 &     0.18\\
\hline
\end{tabular}
\end{flushleft}
\label{tabKS}
\end{table*}

\subsection{The influence of errors}
\label{errors}

\begin{figure}[ht!]
\resizebox{\hsize}{!}{\includegraphics{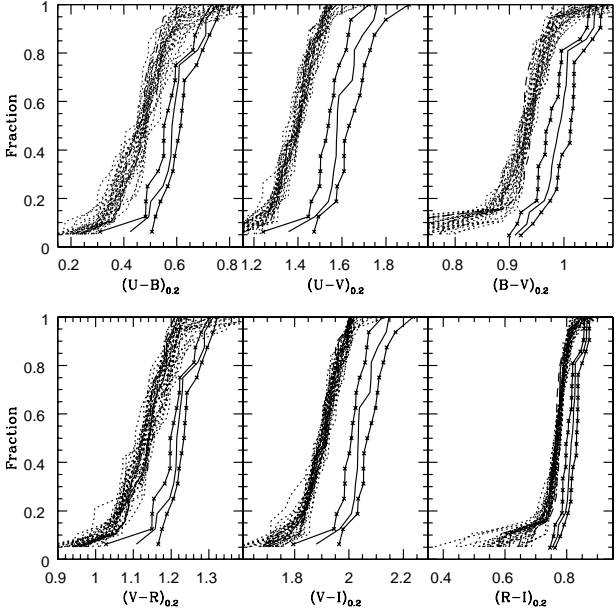}}
\caption[Color gradients]{The influence of statistical and systematic 
errors on the
cumulative distributions of the observed and predicted central colors 
for the LS (upper) and
the DS (lower) galaxies. For the DS galaxies and for each color we
show 30 distributions (dotted lines) obtained by bootstrapping the
measured colors and modeling their evolution (see text for
explanation). The thick lines show the measured distributions. The
dashed thick lines show the effects of systematic errors. For the LS
galaxies we show the measured and evolved distributions taking into
account the systematic errors (lines with crosses).}
\label{figerrmodcol}
\end{figure}

\begin{figure}[ht!]
\resizebox{\hsize}{!}{\includegraphics{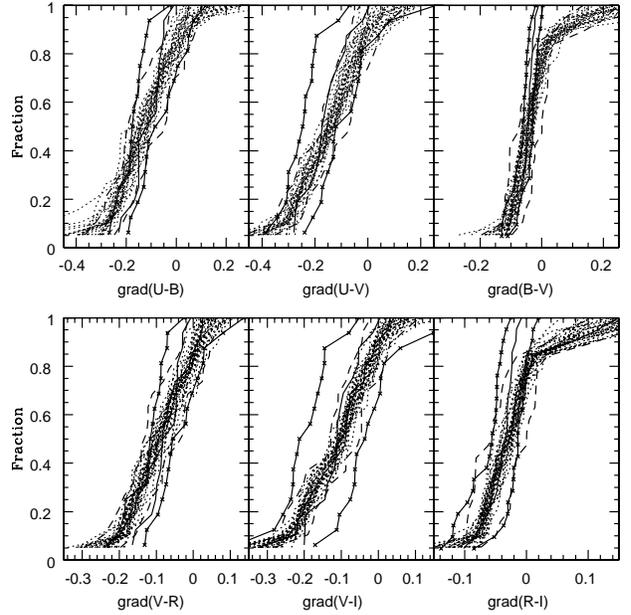}}
\caption[Color gradients]
{The influence of statistical and systematic errors on the
cumulative distributions of the observed and predicted color gradients
for the LS (upper) and the DS (lower) galaxies. For the DS galaxies
and for each color we
show 30 distributions (dotted lines) obtained by bootstrapping the
measured colors and modeling their evolution (see text for
explanation). The thick lines show the measured distributions. The
dashed thick lines show the effects of systematic errors. For the LS
galaxies we show the measured and evolved distributions taking into
account the systematic errors (lines with crosses).}
\label{figerrmod}
\end{figure}

In the previous sections we pointed out that some of the problems
recognized in the modeling (the difference in the median central
colors of the LS and DS samples, the presence of positive gradients at
$z\approx 0.4$, the differences in the observed and predicted width of
the gradient distributions) might be explained taking into account the
observational errors. Errors are dominated by systematics for the LS
galaxies, and have both a statistical and a systematic component for
the DS. They affect both the observed 
and the predicted distributions. We estimate the effect of statistical
errors
on the observed distributions by bootstrapping the measured colors or color
gradients with their statistical errors, assumed to be gaussianly
distributed. 
The estimation of the effect on the predicted color and color gradient
distributions requires the choice of a model. For simplicity, we focus
on the m{\bf M} Case with present-day age of 10 Gyr. 
We bootstrap the measured colors at 0.2 and 2 $R_e$ with their
statistical errors at $z\approx 0.4$, 
assumed to be gaussianly distributed. We compute
the metallicities implied by the SSP models, predict the expected
colors at z=0, and derive the color
gradients and their cumulative distributions. Figs. \ref{figerrmodcol}
and \ref{figerrmod} show the result of 30 such bootstraps.  
The typical width of the obtained measured and predicted 
color distribution clouds is
$\approx 0.02$ mag. As discussed in Sect. \ref{cl0952}, 
the statistical errors dominate over the systematic effects.
The typical width of the obtained measured and predicted distributions 
clouds of the color gradients is 
$\approx 0.02-0.04$ mag/dex. In this case
the systematic effects bracket the statistical variations.

Fig. \ref{figerrmodcol} shows that  measurement
errors alone are insufficient to explain the difference in central
color between
the LS and DS samples, because the simulated and observed
distributions at both z=0 and $z\approx0.4$ are separated even when
errors on both sides are considered. A larger look-back
time (8 Gyr) produces a good overlap.

The inspection of Figure \ref{figerrmod} shows
first that the results described in Sect. \ref{distgrad} are robust
against the errors. Cases a{\bf A} and m{\bf A} with a central present
age of  10 Gyr
are ruled out with high significance because the large tail
of steep negative gradients that they produce at $z\approx 0.4$ is not 
present in any of the distributions with errors. As a second remark,
errors help explaining the discrepancies between the observed and
modeled distributions. Complete overlap at both z=0 and $z\approx 0.4$
is achieved between the observed and modeled distributions of the U-B
and $V-R$ gradients. The agreement for the $U-V$ and $V-I$ gradient
distributions is improved, with most of the differences in the widths
accounted by the errors. The 10\% tail of distant galaxies with the
largest positive R-I gradients (objects 211 and 312) cannot be
explained by the observational errors. However, the two galaxies are
the smallest of the sample and it can be argued that their gradients
are not well determined (see Sect. \ref{cl0952}). 
Therefore, we conclude that the m{\bf M},
a{\bf M}, and, 
to less extent, m{\bf A}15 cases provide a satisfactory description of the
color gradients of (bright) cluster ellipticals at z=0 and $z\approx
0.4$ following passive evolution.

\subsection{The role of dust}
\label{dust}

Wise \& Silva (\cite{WS96}) examined the possibility that color
gradients in elliptical galaxies are caused entirely by dust effects. 
They fit the color profiles of the 
LS galaxies considered here adopting different radial distributions
and total mass of the dust component. Their best fitting models are
characterized by a distributions $\propto r^{-1}$ and dust masses 
much higher than those directly determined from the IRAS data. 
The ratios of these masses
are in the range 3.6-307, only NGC 1399 having ratio of
order unity. 
Therefore we are confident that the radial variation of the 
stellar populations plays a major role in driving 
the color gradients.  Nevertheless, part of the gradient may be due to dust.
We have thus tested the hypothesis that 50\% of the color gradients 
observed are 
caused by dust, to verify how robust our conclusions of
Sect. \ref{distgrad} are. We reduce the gradients by a factor 2
by making the 0.2 $R_e$ colors bluer 
and leaving the 2 $R_e$ as measured. 
As a consequence, for Case a we require central mean ages smaller by 
$\approx 2$ Gyr at z=0 but only shorter by $\approx 0.4$ Gyr at $z\approx
0.4$.  For Case m we require central mean metallicities smaller by
$\approx 0.1$ dex. This does not modify much our conclusions about the
viability of the different mechanisms examined in
Sect. \ref{distgrad}. Case a{\bf A} remains ruled out with high
confidence. A metallicity change (Cases m{\bf M} and a{\bf M}) 
of $\approx 0.1$ dex per radial decade is enough  to match both
local and distant gradient distributions. The Case m{\bf A} with 15 Gyr
central ages and $\approx 2.5$ Gyr younger peripheries  allows a
consistent model of all local and distant objects. The Case m{\bf A} with 10
Gyr central ages is less unprobable than before, but still not acceptable.

These findings can be used to estimate the viability of mixed
scenarios, where combinations of age and metallicity variations are
driving the color gradients. The possibility
that metallicity increases in the outer regions of ellipticals is very
unlikely,
because the age driven model (Case a{\bf A}) cannot explain color
gradients already when the metallicity is kept constant, and works only
with large central ages (Case m{\bf A}15). In this case 0.1 dex increase in
metallicity in the outer parts would require $\approx 2.5$ Gyr
younger ages, enough to attribute ages of the order of $t_{LB}$ to
50\% of the LS galaxies. The possibility that both metallicity and age
decrease outwards can be constrained if the present day central ages of cluster
ellipticals are $\le 10$ Gyr. In this case at most half of the observed color
gradients can be explained by age variations. Scenarios where
metallicity ``overdrives'' the color gradients by compensating 
age increases in the outer regions of the galaxies are also
possible. Approximately 0.1 dex additional decrease in metallicity is needed to
compensate 2.5 Gyr increase in age every radial decade. 

\section{Conclusions}
\label{conclusions}

We presented the surface brightness profiles of the 20 brightest
early-type galaxies of CL0949 at redshift $z$=0.35-0.38 from HST WF2
frames taken with the filters F555W, F675W, F814W. We determined the
color profiles $(V-R)(r)$, $(V-I)(r)$, and $(R-I)(r)$, and fit
logarithmic gradients in the range $-0.2$ to 0.1 mag per decade.
These values are similar to what is found locally for the colors $(U-B)$,
$(U-V)$, $(B-V)$ which approximately match the $(V-R)$, $(V-I)$,
$(R-I)$ at
redshift $\approx 0.4$.  We analyzed the results with up to date
stellar population models, exploring whether the following mechanisms
are able to explain the colors of cluster ellipticals and their
gradients assuming passive evolution between redshift z=0 and 0.4. {\bf
Case} a{\bf A:}  The central colors and
their gradients are due to age differences. Galaxies have all solar
metallicity. {\bf Case} m{\bf M}:  The central colors and
their gradients are due to metallicity differences. Galaxies
have all the same age (10 or 15 Gyr at present). {\bf Case} a{\bf M:}
The color gradients are due to metallicity.
Galaxies have solar metallicity in their centers, the central colors
are defined by the age of the galaxy.
{\bf Case} m{\bf A:} The color gradients are due to age. 
Galaxies have 10-15 Gyr old centers at
present, their central colors are defined by the metallicity of the
galaxy. We reached the following conclusions:

\begin{itemize}
\item The age driven gradient with fixed-metallicity (Case a{\bf A}) 
can be ruled out. It fails
to reproduce the distribution of color gradients of distant 
(predicting too steep negative gradients) and local (predicting too
shallow gradients) galaxies.
\item The metallicity driven gradient with fixed-age (Case m{\bf M}) 
is viable for both galaxy ages
assumed (10 or 15 Gyr at the present day). A metallicity gradient
of $\approx 0.2$ dex per radial decade is needed to explain the
observed color gradients of local and distant galaxies. The gradient
distributions at z=0 and 0.4 can be matched well by taking into account the
observational errors.
\item The metallicity driven gradients model assigning a fixed (solar)
metallicity to the centers (Case a{\bf M}) works
also well, being very similar to case m{\bf M} for a 10 Gyr age. 
\item The age driven gradient model assigning a fixed age to the
centers (Case m{\bf A}) does not work if the 
present day central age is smaller than or equal to 10 Gyr. In this case
too steep negative gradients at $z\approx 0.4$ are produced and 
furthermore the modeling of some $z=0$ galaxies is not possible, 
because it would require ages smaller than $t_{LB}$. A
central age of 15 Gyr allows the modeling of most galaxies with an
age gradient of $\approx 4$ Gyr per radial decade. 
\item The central colors of local (bright) cluster galaxies imply
central ages of $\approx 10$ Gyr (if solar metallicity is assumed) or
slightly subsolar ([Fe/H]$\approx -0.1$) metallicity, if a central age
of 15 Gyr is assumed.
\item  With $t_{LB}=4.53$ local galaxies appear too red
at $z\approx 0.4$ and distant galaxies appear too blue at z=0.
Possible reasons are the uncertainties in the galaxy redshifts, 
in the calibration to the Johnson-Cousins system at $z\approx 0.4$ and
 the U band at $z=0$. 
The distributions of the central colors of local and distant
ellipticals can also be matched by passive evolution of simple stellar
populations if large look-back times ($\approx 8$ Gyr) are assumed. 
\item The results described are obtained consistently in the three
pairs of colors ($U-B$ and $V-R$,~$U-V$ and $V-I$, $B-V$ and $R-I$).
\item The conclusions reached above are still valid if the measured
gradients are reduced by a factor 2 to simulate the possible presence
of a screen of dust. A 0.1 dex reduction of metallicity per radial
decade would suffice in Cases m{\bf M} and a{\bf M} to explain the
data. Case m{\bf A}
with 15 Gyr present day central age and $\approx 2.5$ Gyr younger
outer regions would also be a possible explanation.
\end{itemize}

Bright cluster ellipticals appear 10-15 Gyr old at z=0 
whatever the modeling assumptions. This is
an important constraint that models of galaxy formation must take
into account. 

The present results do not allow per se to discriminate whether
metallicity or age are generating the color-magnitude relation of
cluster ellipticals. Both m{\bf M} and a{\bf M} cases provide a 
reasonable model of
the color gradients observed locally and at $z\approx 0.4$ down to the
magnitude limit of our sample ($B\approx -21$). However,
the metallicity scenario is favored when the CM relation down to
fainter magnitudes is considered (Kodama and Arimoto \cite{KA97}).

The presence of a metallicity gradient in galaxies (m{\bf M} or 
a{\bf M} case) is our favored mechanism for the origin of color 
gradients, since a
fraction of galaxies cannot be modeled under m{\bf A}15. 
Tamura et al. (\cite{Tetal2000a}) and Tamura and Ohta (\cite{Tetal2000b}) 
reach the same conclusion from the
analysis of the color gradients of ellipticals in the Hubble Deep
Field North and in clusters. This indicates that our result does
 not depend on our particular choice of models.

The mild center to
periphery metallicity difference (-0.2 dex per radial decade)
reiterates the failure of classical monolithic collapse models with
Salpeter IMF, which predict much stronger gradients ($\approx -0.5$
dex between center and $R_e$, Carlberg \cite{C84}). In hierarchical
galaxy formation scenarios, originally present 
gradients are diluted by the merging process (White \cite{W80}).
This goes in the direction of accounting for the observations, but 
quantitative predictions are still
lacking. Passive evolution seems adequate to explain the color
gradients at both z=0 and $z\approx 0.4$, adding to the evidence in
its favor cumulated in recent years (see Sect. \ref{introduction}). The
real origin of color gradients might actually reside in a combination
of age and metallicity effects. However, we can rather safely exclude that 
metallicity increases in the outer parts of galaxies, because age
variations cannot explain the color gradients when a constant
metallicity is assumed (Case a{\bf A}) and are just viable when
large central ages are adopted (Case m{\bf A}15). We can also exclude
that more than half of the observed color gradients is driven by age
variations, if the present day central ages of cluster
ellipticals are $\le 10$ Gyr. Scenarios where
metallicity ``overdrives'' the color gradients by compensating 
age increases in the outer regions of the galaxies are also
possible. Approximately 0.1 dex additional decrease in metallicity is needed to
compensate 2.5 Gyr increase in age every radial decade. 

In principle the use of both the H$\beta$ (more sensitive to age
variations) and the Mg$_2$ (more sensitive to metallicity variations)
line indices could allow to disentangle age and metallicity effects 
(Worthey \cite{W94}). Following this idea,  
studies of the H$\beta$ and Mg$_2$ line index gradients in field 
(Gonz\'alez \cite{G93}) and
in Coma cluster ellipticals (Mehlert et al. \cite{M00a}) find on the mean
that the outer regions of early-type galaxies are metal poorer (0.1
dex per decade in radius) and slightly older (0.04 dex per decade in
radius) than the inner parts (Mehlert et al. \cite{M00b}, see also
Bressan et al. \cite{BCT96} and Kobayashi \& Arimoto \cite{KA99}). Our
modeling of color gradients requires steeper metallicity gradients
($\approx 0.3$ dex per decade) to compensate for these kinds of older
stellar populations in the outer regions of galaxies. Note, however,
that the modeling of line indices is uncertain (Maraston, Greggio \&
Thomas \cite{MGT99}). Moreover, the possible presence of old, metal poor
stellar populations complicates the interpretation of Balmer
lines as age indicators (Maraston \& Thomas \cite{MT99}). 

It is not clear that stronger conclusions on the origin of color
gradients of cluster ellipticals could be achieved looking at more
distant clusters. On the one hand, even the resolution of HST might not be
enough to allow their reliable determination down to the same limiting
magnitude at larger redshifts. On the other hand, we may likely find
traces of residual star formation (e.g., Poggianti et al. \cite{Pog98})
that make the interpretation of the evolution impossible, adding too
many free parameters to the problem. Dealing with lower redshift
cluster might also not be very informative, because the expected
evolution is small. Improved constraints can instead be
obtained by enlarging the sample of local and $z\approx 0.4$
cluster galaxies, probing the brighter and fainter end of the luminosity
function. In particular, fainter ellipticals have bluer colors 
which might require too young ages and rule out the age driven 
origin of color gradients completely. 

\begin{acknowledgements}
This research was supported by the Sonderforschungsbereich
375. Some image reduction was
done using the MIDAS and/or IRAF/STSDAS packages. 
IRAF is distributed by the National Optical
Astronomy Observatories, which are operated by the Association of
Universities for Research in Astronomy, Inc., under cooperative
agreement with the National Science Foundation. 
We thank S. Cassisi
for having provided us the complete set of stellar tracks with
half-solar metallicity. Finally, we thank the referee, R. Peletier,
for his valuable comments.  
\end{acknowledgements}


\begin{thebibliography}{}

\bibitem[1999]{AEFTG99}
Abraham, R.G., Ellis, R.S., Fabian, A.C., Tanvir, N.R., Glazebrook,
K., 1999,  { MNRAS} { 303}, 641

\bibitem[1986]{AY86}
Arimoto, N., Yoshii, Y., 1986, A\&A, 164, 260

\bibitem[1996]{Betal96}
Baugh, C.M., Cole, S., Frenk, C.S., 1996,  { MNRAS}, {283}, 1361

\bibitem[1987]{BM87}
Bender, R., M\"ollenhoff, C., 1987,  {A\&A} {177}, 71

\bibitem[1996]{BZB96}
Bender, R., Ziegler, B., and Bruzual, G., 1996,
 { ApJ} { 463}, L51

\bibitem[1998]{BSZBBGH97}
Bender, R., Saglia, R.~P., Ziegler, B., Belloni, P., Bruzual, G., Greggio, L.,
  and Hopp, U., 1998,  { ApJ} { 493}, 529

\bibitem[1998]{Betal98}
Bernardi, M., Renzini, A., da Costa, L.N., Wegner, G., Alonso, M.V.,
Pellegrini, P.S., Rit\'e, C., Willmer, C.N.A., 1998, 
{ ApJL}, { 508}, 143

\bibitem[1996]{BA96}
Bertin, E., Arnouts, S., 1996,  { A\&AS} { 117}, 393 

\bibitem[1979]{B79}
Bessel, M.S., 1979, PASP, 91, 589

\bibitem[1997]{bono97}
Bono, G., Caputo, F., Cassisi, S., Castellani, V. \& Marconi, M.,
1997,  { ApJ}, { 489}, 822

\bibitem[1992]{Betal92}
Bower R.G., Lucey J.R., Ellis R.S., 1992,  { MNRAS}, { 254}, 601

\bibitem[1996]{BCT96}
Bressan, A., Chiosi, C., Tantalo, R., 1996, A\&A, 311, 425

\bibitem[1996]{BC99}
Bruzual, G., Charlot, S., 1996, in preparation

\bibitem[1984]{BH84}
Burstein, D., Heiles, C., 1984,  {ApJS} { 54}, 33

\bibitem[1978]{B78}
Buser, R., 1978, A\&A, 62, 411

\bibitem[1984]{C84}
Carlberg R.G., 1984,  {ApJ} { 286}, {403}

\bibitem[1993]{C93}
Carollo, M., Danziger, I.J., Buson, L., 1993,  { MNRAS} { 265}, 553

\bibitem[1999]{C99}
Castelli, F., 1999, A\&A, 346, 564

\bibitem[1988]{CBB88}
Chiosi, C., Bertelli, G., Bressan, A., 1988, A\&A, 196, 84

\bibitem[1999]{Cetal99}
Colless, M., Burstein, D., Davies, R.L., McMahan, R.K., Saglia, R.P., 
Wegner, G., 1999,  { MNRAS}, { 303}, 813

\bibitem[1998]{Cou98}
Couch, W.J., Barger, A.J., Smail, I. Ellis, R.S., Sharples, R.M.,
1998, ApJ 497, 188

\bibitem[1993]{D93}
Davies R.L., Sadler E.M., Peletier R.F., 1993,  { MNRAS},
{ 262}, 650

\bibitem[1996]{dJ96}
de Jong, R.S., 1996,  { A\&A} { 313}, 377

\bibitem[1961]{dV61}
de Vaucouleurs, G., 1961,  { ApJS} { 5}, 223

\bibitem[1992]{DG92}
Dressler, A., Gunn, J.E., 1992,  { ApJS} { 78}, 1

\bibitem[1997]{Eetal97}
Ellis, R.S., Smail, I., Dressler, A., Couch, W.J., 
Oemler, A. Jr., Butcher, H., Sharples, R.M., 1997, ApJ, 483, 582

\bibitem[1977]{F77}
Faber, S.M., 1977,  in {The Evolution of Galaxies and Stellar
Populations} {eds. B. Tinsley, R.B. Larson} (New Haven, Yale
Univ. Obs.), 157

\bibitem[1989]{FIH89}
Franx, M., Illingworth, G., Heckman, T., 1989,  { ApJ}, { 344}, 613

\bibitem[1976]{FPR76}
Fusi Pecci, F., Renzini, A., 1976, A\&A, 46, 447

\bibitem[1993]{G93}
Gonz\'alez, J.J., 1993,  {PhD Thesis} { University of
California}, Santa Cruz

\bibitem[1994]{GHJNDV94}
Goudfrooij, P., Hansen, L., Joergensen, H.E., Noergaard-Nielsen, H.U.,
De Jong, T., and Van Den Hoek, L.B., 1994,  { A\&AS} { 104}, 179

\bibitem[1995]{GdJ95}
Goudfrooij, P., de Jong, T., 1995,  { A\&A} { 298}, 784

\bibitem[1995]{HBCHTWW95}
{Holtzman}, J.~A., {Burrows}, C.~J., {Casertano}, S., {Hester}, J.~J.,
  {Trauger}, J.~T., {Watson}, A.~M., and {Worthey}, G., 1995,
 { PASP} { 107}, 1065

\bibitem[1996]{JFK96}
J{\o}rgensen, I., Franx, M., and Kj{\ae}rgaard, P., 1996,
 {MNRAS} { 280}, 167

\bibitem[1996]{K96}
Kauffmann, G., 1996,  { MNRAS}, { 281}, 487

\bibitem[1997]{KDFIF97}
Kelson, D.~D., van Dokkum, P. G., Franx, M., Illingworth, G.~D., and Fabricant,
  D., 1997,  { ApJ} { 478}, L13

\bibitem[1999]{KA99}
Kobayashi, C., Arimoto, N., 1999, ApJ, 527, 573

\bibitem[1997]{KA97}
Kodama, T., Arimoto, N., 1997,  {A\&A} { 320}, 41

\bibitem[1979]{K79}
Kurucz, R., 1979, ApJS, 40, 1

\bibitem[1993]{LC93}
Lacey, C., Cole, S., 1993,  { MNRAS} {262}, 627

\bibitem[1974]{L74}
Larson, R.B., 1974,  { MNRAS} { 173}, 671

\bibitem[1998]{LCB98}
Lejeune, T., Cuisinier, F., Buser, R., 1998, A\&AS 130, 65

\bibitem[1998]{Ma98}
Maraston, C., 1998,  { MNRAS} { 300}, 872

\bibitem[2000]{Ma99}
Maraston, C., 2000,  in preparation

\bibitem[1999]{MGT99}
Maraston, C., Greggio, L., Thomas, D., 1999,  Ap\&SS in press

\bibitem[2000]{MT99}
Maraston, C., Thomas, D., 2000, ApJ in press, astro-ph/0004145

\bibitem[1998]{M98}
Mehlert, D., Saglia, R.P., Bender, R., Wegner, G., 1998, 
{ A\&A}, { 332}, 33

\bibitem[2000a]{M00a}
Mehlert, D., Saglia, R.P., Bender, R., Wegner, G., 2000a, 
{ A\&AS}, 141, 449 

\bibitem[2000b]{M00b}
Mehlert, D., Saglia, R.P., Bender, R., Wegner, G., 2000b, 
in preparation

\bibitem[1990]{PDIDC90}
Peletier, R.F., Davies, R.L., Illingworth, G.D., Davis, L.E., and Cawson, M.,
1990,  { AJ} {100}, 1091 


\bibitem[1998]{Pog98}
Poggianti, B.M., Smail, I. Dressler, A., Couch, W. J., Barger, A.J.,
Butcher, H. Ellis, R.S., Oemler, A. Jr., 1998, ApJ 518, 576

\bibitem[1975]{R75}
Reimers, D., 1975, Mem. Soc. R. Sci. Liege, Ser. 6, Vol. 8, 369

\bibitem[1986]{RB86}
Renzini, A. and Buzzoni, A., 1986 in Spectral Evolution of Galaxies, C.
Chiosi and A. Renzini (eds.), Dordrecht, Reidel, 195

\bibitem[1997a]{SBBBCDMW97a} 
Saglia, R.P., Burstein, D., Baggley, G., Bertschinger, E.,
Colless, M.M., Davies, R.L., McMahan, R.K., Wegner, G., 1997a, 
 { MNRAS} {292}, 499

\bibitem[1997b]{SBBBCDMW97b}
{Saglia}, R.~P., {Bertschinger}, E., {Baggley}, G., {Burstein}, D., {Colless},
  M., {Davies}, R.~L., {McMahan}, Robert~K., J., and {Wegner}, G., 1997b,
 { ApJS} { 109}, 79

\bibitem[1998]{SFD98}
Schlegel, D.J., Finkbeiner, D.P., Davis, M., 1998, ApJ 500, 525

\bibitem[1998]{SSBHBZ98}
Seitz, S., Saglia, R.P., Bender, R., Hopp, U., Belloni, P., Ziegler,
B., 1998,  { MNRAS} { 298}, 945

\bibitem[1998]{SED98}
Stanford, S.A., Eisenhardt, P.R., Dickinson, M., 1998, 
{ ApJ} { 492}, 461

\bibitem[2000]{Tetal2000a}
Tamura, N., Kobayashi, C., Arimoto, N., Kodama, T., Ohta, K., 2000,
AJ, 119, 2134

\bibitem[2000]{Tetal2000b}
Tamura, N., Ohra, K., 2000, AJ, in press (astro-ph/0004221)

\bibitem[1996]{Tetal96}
Tantalo, R., Chiosi, C., Bressan, A., Fagotto, F., 1996, A\&A, 311, 361

\bibitem[1999]{TGB99}
Thomas, D., Greggio, L., Bender, R., 1999, 
 { MNRAS} { 302}, 537

\bibitem[1999]{T99}
Treu, T., Stiavelli, M., Casertano, S., M\o ller, P., Bertin, G.,
1999,  { MNRAS} { 308}, 1037

\bibitem[1998]{DFKI98}
van Dokkum, P. G., Franx, M., Kelson, D.~D., and Illingworth, G.~D., 1998,
 { ApJ} { 504}, L53

\bibitem[1996]{Vetal96}
Vazdekis, A., Casuso, E., Peletier, R.F., Beckman, J.E., 1996, ApJSS,
106, 307

\bibitem[1978]{WR78}
White S.D.M., Rees, M.J., 1978, MNRAS, 183, 341

\bibitem[1980]{W80}
White S.D.M., 1980,  { MNRAS} { 191}, 1

\bibitem[1996]{WS96}
Wise, M.W., Silva, D.R., 1996,  { ApJ} { 461}, 155

\bibitem[1992]{WFG92}
Worthey, G., Faber, S.M., Gonz\'alez, J.J., 1992,  { ApJ} { 398}, 69

\bibitem[1994]{W94}
Worthey, G., 1994,  { ApJS} { 95}, 107

\bibitem[1997]{ZB97}
Ziegler, B.~L. and Bender, R., 1997,
 { MNRAS} { 291}, 527

\bibitem[1999]{ZSBBGS99}
Ziegler, B.L., Saglia, R.P., Bender, R.,
Belloni, P., Greggio, L., and Seitz, S., 1999,  { A\&A} { 346}, 13
\end{thebibliography}
\end{document}